\documentclass[review, times, 12pt]{elsarticle}

\usepackage[margin=1in]{geometry}
\usepackage{setspace}
\usepackage{amssymb}
\usepackage{amsfonts}
\usepackage{amsmath}
\usepackage{amsthm}
\usepackage{txfonts}
\usepackage{textcomp}
\usepackage{graphicx}
\usepackage[english]{babel}
\usepackage{epsfig}
\usepackage{subcaption}
\usepackage{xcolor}
\usepackage{color,soul}
\usepackage{multirow}
\usepackage{makecell}
\usepackage{tabularx}
\usepackage{todonotes}
\usepackage{booktabs}
\usepackage{pifont}
\usepackage{multirow}
\usepackage{diagbox}
\usepackage{mathtools}
\usepackage{lipsum}
\usepackage{wrapfig}
\usepackage{soul}
\usepackage{threeparttable}
\usepackage{siunitx}
\usepackage{hyperref} 

\newtheorem{hypothesis}{Hypothesis}

\newcolumntype{C}[1]{>{\centering\arraybackslash}m{#1}}

\journal{Additive Manufacturing}

\begin{document}
\begin{frontmatter}
\title{A reduced-order analysis for multi-reflection-driven laser absorptance in vaporization-induced cavities}

\author[label1,label2,label3]{{Haolin~Liu}\corref{cor1}}
\author[label4]{{Haoran~Shi}}
\author[label1]{{Xuzhe~Zeng}}
\author[label5]{{Brian~J.~Simonds}}
\author[label4]{{Wenda~Tan}\corref{cor1}}
\author[label2,label3]{{Anthony~D.~Rollett}\corref{cor1}}
\author[label1,label3]{{Levent~Burak~Kara}\corref{cor1}}

\address[label1]{Department of Mechanical Engineering, Carnegie Mellon University, Pittsburgh, Pennsylvania, USA}

\address[label2]{Department of Materials Science and Engineering, Carnegie Mellon University, Pittsburgh, Pennsylvania, USA}

\address[label3]{NextManufacturing Center, Carnegie Mellon University, Pittsburgh, Pennsylvania, USA}

\address[label4]{Department of Mechanical Engineering, University of Michigan, Ann Arbor, Michigan, USA}

\address[label5]{National Institute of Standards and Technology, 100 Bureau Drive, Gaithersburg, Maryland, USA}

\cortext[cor1]{Corresponding authors. \\\hspace*{1.6em} E-mail address: haolinl@mit.edu; wendatan@umich.edu; \{rollett, lkara\}@andrew.cmu.edu}

\begin{abstract}
Multiple internal reflections enhance laser energy coupling within vaporization-induced cavities, yet quantitatively connecting finite-beam ray histories, cavity geometry, and total absorptance conventionally requires full ray tracing. Here, we develop a reduced-order analysis in which repeated laser--wall interactions produce a Beer--Lambert-type attenuation relation governed by an effective per-reflection optical depth~(EPROD). A power-weighted finite-beam formulation defines the absorptance-equivalent interaction count, $N_{\mathrm{eff}}$, exactly under the constant effective single-interaction absorptance approximation, while a cumulant expansion separates the leading power-weighted mean interaction count, $\overline{N}_w$, from systematic corrections associated with ray-path heterogeneity. Because $N_{\mathrm{eff}}$ depends explicitly on EPROD, $\overline{N}_w$ is retained as the EPROD-independent leading statistic for identifying the geometric interaction-accumulation scale. Across nine ray-resolved simulation groups comprising three alloys and three laser spot diameters, the unit-intercept relation $\overline{N}_w=1+cd$ captures the dominant depth dependence with $R^2=0.7824$--$0.9843$. At fixed spot diameter, the cross-alloy variation in $c$ is only $3.4\%$--$6.3\%$, whereas increasing the spot diameter from $77$ to $123~\mu\mathrm{m}$ reduces $c$ by $15\%$--$29\%$. With $c$ fixed from the ray statistics and only $b$ fitted, the resulting absorptance--depth relation achieves a median MAPE of $2.61\%$, while $b$ remains within the narrow range $0.9089$--$0.9337$. An independent synchronized Ti-64 experiment under nearly matched beam geometry yields $c=0.0168~\mu\mathrm{m}^{-1}$ and $b=0.9059$, each within approximately $2.4\%$ of the corresponding simulation value. The framework therefore establishes a computationally efficient and physically interpretable factorization of cavity geometry, material optical attenuation, and global absorptance, supporting rapid energy-coupling prediction and absorptance-based keyhole characterization.
\end{abstract}

\begin{graphicalabstract}
\label{sec:graphical}
\begin{figure*}[!ht]
\center{\includegraphics[width=0.8\linewidth]
{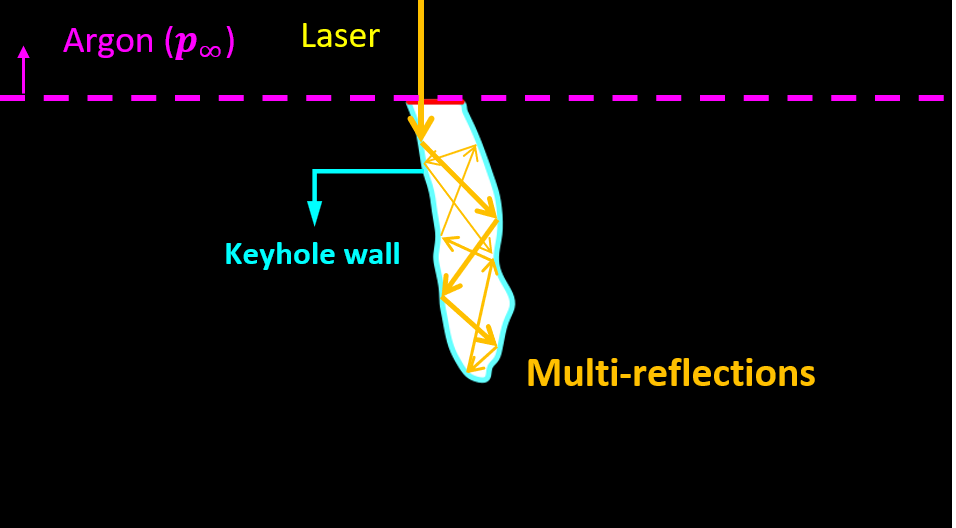}
}
\end{figure*}
\end{graphicalabstract}

\begin{highlights}
\item Finite-beam ray histories are reduced to an absorptance-equivalent interaction count with systematic corrections for ray-path heterogeneity.
\item The EPROD-independent power-weighted mean interaction count admits a leading-order linear depth representation across nine material--beam conditions.
\item The interaction-density coefficient varies weakly across alloys but decreases systematically with increasing laser-beam spot size.
\item The resulting depth--absorptance closure predicts the nine simulation conditions with a median MAPE of $2.61\%$.
\item Independent synchrotron experiment recovers both fitted closure parameters within approximately $2.4\%$ of the matched simulation.
\end{highlights}


\end{frontmatter}


\section{Introduction}
\label{sec:intro}
\begin{figure}[!ht]
\centering
\includegraphics[width=0.8\linewidth]{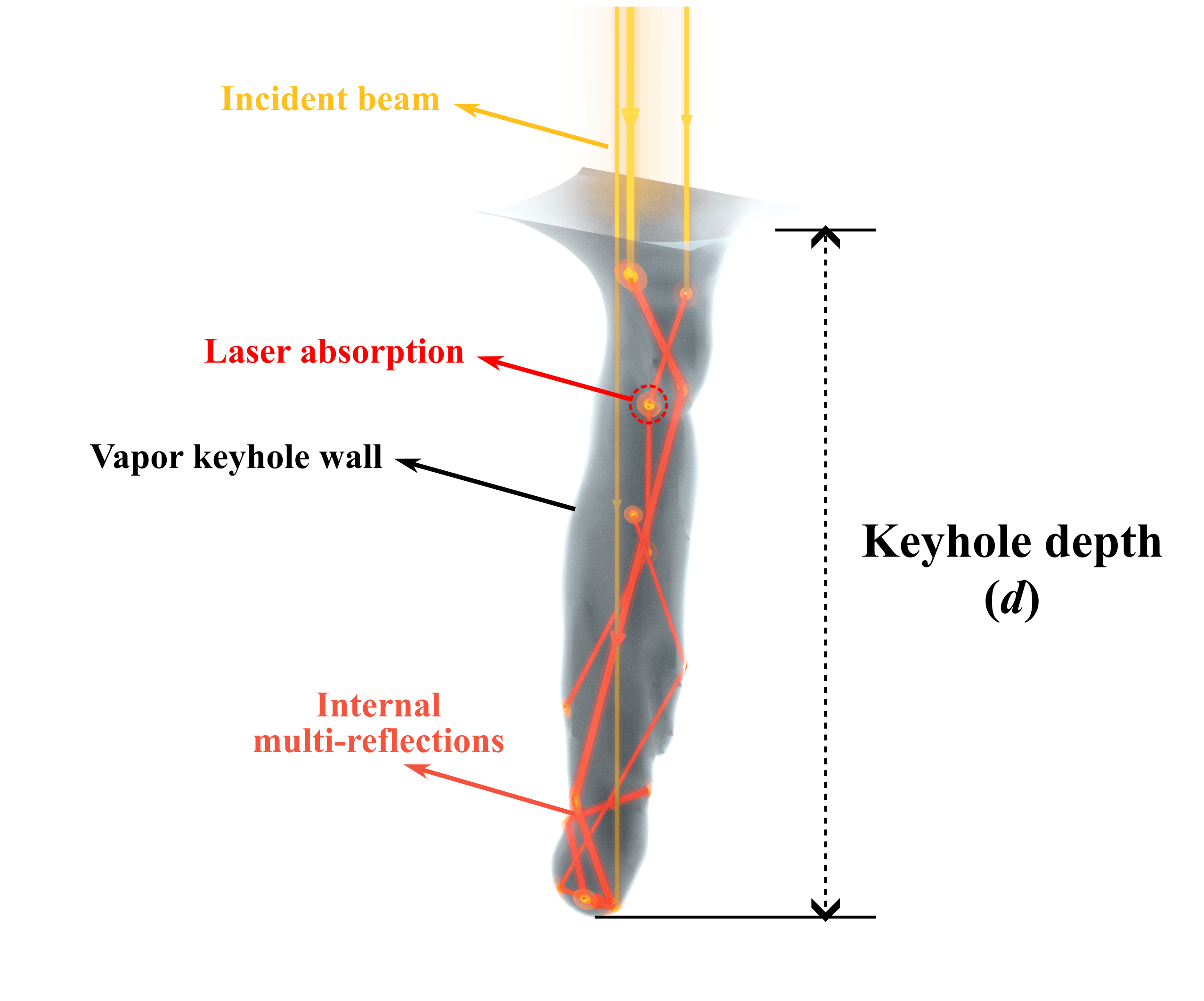}
\caption{Multi-reflection-driven laser absorptance within a vaporization-induced cavity (\textit{keyhole}). }
\label{fig:laser_multi_reflection}
\end{figure}

In laser-based metal processing (LMP), including laser welding and powder-bed fusion (PBF) additive manufacturing (AM), concentrated laser irradiation locally melts metal, which subsequently resolidifies to form joints or dense components~\cite{Fabbro2005Study,king2015laser}. At sufficiently high energy densities, the melt-pool surface can surpass the alloy's vaporization temperature, resulting in intense evaporation~\cite{Fabbro2005Study,khairallah2016laser}. As illustrated in Fig.\ref{fig:laser_multi_reflection}, the associated recoil pressure depresses the liquid surface and forms a vapor cavity, commonly termed a \textit{keyhole} in LMP and laser-PBF communities, into which the incident laser beam propagates and undergoes repeated interactions with the cavity walls~\cite{cunningham2019keyhole,zhang2013direct}. These internal reflections enhance energy trapping, thereby increasing the total absorptance and promoting deeper penetration~\cite{king2015laser,kaplan2012fresnel,martin2019dynamics, deisenroth2026illuminating}. The coupled feedback among energy deposition, multiple reflections, and the evolving cavity geometry governs keyhole development and its associated dynamics~\cite{khairallah2016laser,zhao2022laser}. A compact quantitative relationship between keyhole geometry and total absorptance would therefore provide a valuable physical closure for process modeling, energy-deposition prediction, and \textit{in~situ} characterization. 

Experimental studies have progressively revealed this geometry--absorptance coupling. Calorimetric measurements by Trapp~\textit{et~al.}~\cite{trapp2017situ} showed that effective laser absorptance varies substantially with processing conditions and approaches a plateau after forming a recoil-pressure-induced depression. Simonds~\textit{et~al.}~\cite{simonds2018time} subsequently resolved sub-microsecond absorptance evolution during laser spot welding and identified distinct changes associated with melting and keyhole formation. Across {Ti--6Al--4V}~(Ti-64), Alloy~718, and stainless steel 316 (SS316), Ye~\textit{et~al.}~\cite{ye2019energy} related measured absorptance and melt-pool depth to normalized processing variables and reported an asymptotic, approximately exponential absorptance trend with respect to laser power. More directly, Allen~\textit{et~al.}~\cite{allen2020energy} and Deisenroth~\textit{et~al.}~\cite{deisenroth2026illuminating} both employed real-time measurement approaches to establish correlations between keyhole dynamics and laser energy absorption supported by ray-tracing analysis. These studies revealed a strong positive correlation between keyhole depth and laser absorptance in the keyhole regime and associated stepwise changes in absorptance with discrete changes in reflection count. Collectively, these observations identify reflection accumulation as a key intermediate process linking cavity geometry to energy coupling, but they do not, by themselves, provide a compact physical interpretation relating these quantities across processing conditions. 

On the modeling side, ray-tracing simulations have been widely used to explicitly resolve the reflection process. These approaches combine {Fresnel} absorption with geometric ray propagation in transient free-surface simulations~\cite{cho2006implementation}, laser-PBF melt-pool models~\cite{khairallah2016laser}, and multiphysics frameworks assessed against synchrotron measurements~\cite{kouraytem2019effect}. Reconstructed three-dimensional keyhole geometries have also been used to calculate detailed ray histories and spatial absorption distributions~\cite{jin2012multiple}. Moreover, data-driven methods combined with experimental measurements have also been investigated for estimating laser absorption in laser-PBF~\cite{jiang2024deep}. Although these approaches provide detailed optical information, they require resolving or reconstructing the full cavity geometry and performing numerous ray--surface interaction calculations. Such processes entail substantial computational expense, motivating the development of a tractable description that reduces this burden. In the heat-transfer model of Zhao and DebRoy~\cite{zhao2003macroporosity} and its subsequent application by Rai~\textit{et~al.}~\cite{rai2007heat}, the locally absorbed intensity was represented analytically by a geometric series expression, with an average reflection count inferred from the mean keyhole-wall angle. Although this formulation demonstrates that cumulative reflection absorption can be treated theoretically, it was assessed only as one component of complete welding simulation schemes through comparisons with macroscopic weld geometries. Together, these studies establish the physical importance of cumulative reflection absorption and the value of ray-resolved modeling. They also create an opportunity for an intermediate description that retains the governing reflection mechanism while reducing the complete ray histories and evolving cavity geometry to a small number of physically interpretable quantities. 

In this work, we develop a reduced-order analysis that connects finite-beam interaction statistics, vaporization-cavity depth, and total laser absorptance. Under a constant effective single-interaction absorptance, the beam-level response can be represented exactly through an absorptance-equivalent interaction count and a Beer--Lambert-type attenuation relation. A cumulant expansion then separates the leading power-weighted mean interaction count, $\overline{N}_w$, from corrections associated with ray-path heterogeneity. Although $N_{\mathrm{eff}}$ provides the absorptance-equivalent statistic for a prescribed laser absorptivity, its explicit dependence on the material optical parameter would mix geometric and optical contributions in a fitted depth-scaling coefficient. We therefore test the absorptivity-independent leading relation $\overline{N}_w=1+cd$ across Ti-64, Alloy~718, and SS316 at spot diameters of $77$, $100$, and $123~\mu\mathrm{m}$, while retaining $N_{\mathrm{eff}}$ as a distribution-sensitive reference for evaluating ray-path heterogeneity. For the simulations, $c$ is determined independently from the $\overline{N}_w$ statistics and subsequently held fixed while only the absorptance closure factor $b$ is fitted. A synchronized Ti-64 experiment then provides an independent evaluation in which $c$ and $b$ are jointly inferred from measured absorptance--depth pairs without constraints from the simulation. This factorization separates the leading geometric interaction scale from the material-dependent optical attenuation and establishes a compact route from keyhole geometry to global laser-energy coupling.

\section{Reduced-order analysis of multiple-reflection-driven laser absorptance}
\label{sec:theory}
To begin with, we consider a keyhole with a specific boundary at a given moment of laser--metal interaction. Within this keyhole, where the incident laser keeps ``hitting'' the wall (\textit{i.e.},~the liquid--vapor interface) through multiple reflections, we further assume that: 
\begin{itemize}
    \item The absolute laser single-interaction absorptance when the laser beam impinges on a point of the keyhole wall---denoted as $\varepsilon$---is a material- and laser-dependent constant. 
    \begin{itemize}
    \item This assumption defines a controlled reduced-order representation in which a constant fraction of the incident laser energy is absorbed at each ``hit'' with the alloy's liquid phase. Local variations associated with incidence angle, temperature, polarization, and surface state are replaced by a material-specific effective value of $\varepsilon$; their combined influence on the final depth-based closure is evaluated empirically rather than assumed to vanish.
    \end{itemize}
    \item At each interaction with the keyhole wall, the incident laser energy is partitioned into absorbed and reflected components, with the absorbed component assumed to be consumed primarily by vaporization. Energy losses through other mechanisms are treated as independent of, and therefore separable from, this local absorption--reflection process. 
\end{itemize}

Based on the above assumptions, the total laser absorptance driven by multiple laser--material interactions, denoted as $\eta$, can be expressed as the summation of a geometric series with a common ratio of $1-\varepsilon$:  
\begin{equation}
\label{eqn:eta_1}
    \eta=\varepsilon\sum_{i=1}^N \left(1-\varepsilon\right)^{i-1}=1-\left(1-\varepsilon\right)^N,\:\:N\in\mathbb{Z}^+,\:\:\varepsilon\in\left[0,1\right]
\end{equation}

\noindent where $N$ is the total number of laser--material interactions experienced by the ray, including its initial impingement on the material surface. Eq.~\eqref{eqn:eta_1} has been reported in an early work~\cite{zhao2003macroporosity}. Defining $\tau_r=-\ln{\left(1-\varepsilon\right)}$ and introducing $b$---a phenomenological beam-level closure factor accounting for the residual difference between the idealized constant-EPROD description and the complete optical response\footnote{The closure factor is deliberately defined broadly to encompass unresolved optical effects, including incidence-angle- and temperature-dependent variations in $\varepsilon$, as well as the average distributional difference introduced when the leading mean-count representation is used in the final depth-based model. These contributions are represented collectively by the multiplicative prefactor $b$.}, we have: 
\begin{equation}
\label{eqn:eta_2}
    \eta=b\left[1-\exp{\left(-\tau_r N\right)}\right],\:\:N\in\mathbb{Z}^+,\:\:b\in\left[0,1\right]
\end{equation}

The exponential term in Eq.~\eqref{eqn:eta_2} follows directly from the change-of-base transformation of Eq.~\eqref{eqn:eta_1}, while $b$ extends the idealized relation to a phenomenological beam-level closure. We refer to the transformed dimensionless parameter $\tau_r$ as the \textit{Effective Per-Reflection Optical Depth} (EPROD) of the multi-reflection-driven laser absorption in a keyhole, given that Eq.~\eqref{eqn:eta_2} clearly reveals a Beer--Lambert-type structure. 

Although an individual ray undergoes a discrete, integer number of laser--material interactions, a finite-width laser beam is represented by a continuous spatial distribution of incident power, with rays originating at different locations carrying different power weights and following distinct trajectories within the keyhole. Accordingly, defining a unique beam-level interaction count requires aggregating the local ``ray-level counts'' into a scalar measure using an explicit statistical weighting convention. Let $\Omega_b$ denote the incident laser-beam spot, parameterized by the local in-plane coordinate $\boldsymbol{\xi}$. Let $I(\boldsymbol{\xi})$ denote the incident irradiance with units of $\mathrm{W/m^{2}}$, and let $N(\boldsymbol{\xi})\in\mathbb{Z}^+$ denote the total number of laser--material interactions---including the initial incidence---experienced by the ray entering at $\boldsymbol{\xi}$. The total incident beam power is: 
\begin{equation}
\label{eqn:beam_power}
P_b=\int_{\Omega_b} I(\boldsymbol{\xi})\,\mathrm{d}A
\end{equation}

Following Eq.~\eqref{eqn:eta_2} and the definition of EPROD, we write the resulting total beam absorptance as: 
\begin{equation}
\label{eqn:beam_absorptance}
\eta=\frac{b}{P_b}\int_{\Omega_b}I(\boldsymbol{\xi})\left\{1-\exp\left[-\tau_r N(\boldsymbol{\xi})\right]\right\}\,\mathrm{d}A
\equiv
b\left[1-\exp\left(-\tau_rN_{\mathrm{eff}}\right)\right]
\end{equation}

\noindent where $N_{\mathrm{eff}}$ denotes the \textit{absorptance-equivalent interaction count} and is consequently defined as: 
\begin{equation}
\label{eqn:Neff_continuous}
N_{\mathrm{eff}}=-\frac{1}{\tau_r}\ln\left[\frac{1}{P_b}\int_{\Omega_b}I(\boldsymbol{\xi})\exp\left[-\tau_r N(\boldsymbol{\xi})\right]\,\mathrm{d}A\right]
\end{equation}

Equation~\eqref{eqn:Neff_continuous} is the continuous, incident-power-weighted
counterpart of the LogSumExp transformation and defines the exact beam-level
characteristic interaction count for a given keyhole under the prescribed
constant EPROD. Because it depends on both EPROD and the spatial distribution
of $N(\boldsymbol{\xi})$, $N_{\mathrm{eff}}$ incorporates material optical
properties and morphology-dependent ray trajectories. The corresponding
incident-power-weighted mean and variance for $N(\boldsymbol{\xi})$ are:
\begin{equation}
\label{eqn:weighted_mean_N}
\overline{N}_w=\mathbb{E}_w[N]=\frac{1}{P_b}\int_{\Omega_b}I(\boldsymbol{\xi})N(\boldsymbol{\xi})\,\mathrm{d}A
\end{equation}
\begin{equation}
\label{eqn:weighted_variance_N}
\sigma_{N,w}^{2}=\operatorname{Var}_w[N]=\frac{1}{P_b}\int_{\Omega_b}I(\boldsymbol{\xi})\left[N(\boldsymbol{\xi})-\overline{N}_w\right]^2\,\mathrm{d}A
\end{equation}

\noindent and the cumulant expansion of Eq.~\eqref{eqn:Neff_continuous} is therefore:
\begin{equation}
\label{eqn:cumulant_N}
N_{\mathrm{eff}}=\mathbb{E}_w[N]-\frac{\tau_r}{2!}\operatorname{Var}_w[N]+\frac{\tau_r^{2}}{3!}\kappa_{3,w}-\frac{\tau_r^{3}}{4!}\kappa_{4,w}+\cdots =\overline{N}_w-\frac{\tau_r}{2}\sigma_{N,w}^{2}+\mathcal{R}_{\geq 3}
\end{equation}

\noindent where $\kappa_{m,w}$ denotes the $m$-th incident-power-weighted cumulant, and $\mathcal{R}_{\geq 3}$ collects the third- and higher-order contributions. As defined in Eq.~\eqref{eqn:weighted_mean_N}, $\overline{N}_w$ does not explicitly depend on EPROD and therefore separates the leading geometric contribution of the ray trajectories from the prescribed per-interaction optical attenuation. This absence of explicit EPROD dependence should not be interpreted as complete material independence, because material-dependent cavity evolution may still influence the ray-count distribution through the resulting morphology. Rather, $\overline{N}_w$ provides an EPROD-independent statistic through which the geometric interaction scale can be compared across material--beam conditions.

When the power-weighted distribution of $N(\boldsymbol{\xi})$ is sufficiently concentrated such that the variance and higher-order cumulants are small, Eq.~\eqref{eqn:cumulant_N} gives $N_{\mathrm{eff}}\approx\overline{N}_w$. As the distribution broadens, the negative variance correction lowers $N_{\mathrm{eff}}$ relative to $\overline{N}_w$, with its magnitude modulated by EPROD. More generally, Jensen's inequality gives $N_{\mathrm{eff}}\leq\overline{N}_w$, establishing the mean count as an upper bound. A practical leading-order criterion for the approximation is $\tau_r\sigma_{N,w}^{2}/(2\overline{N}_w)\ll1$, together with small normalized higher-order cumulants; this statistical criterion, rather than depth alone, determines where $N_{\mathrm{eff}}\approx\overline{N}_w$ is justified. Thus, $N_{\mathrm{eff}}$ is retained as the absorptance-equivalent, material-optics-conditioned statistic, whereas $\overline{N}_w$ is used to identify the leading geometric depth scale that is independent of the material's absorptivity.

Given the above analysis, we now propose the following hypotheses: 
\begin{hypothesis}
\label{hyp:linear}
Let $d$ denote the vapor cavity depth. The EPROD-independent power-weighted mean interaction count provides a leading geometric measure of interaction accumulation and increases approximately linearly with $d$ from a unit intercept, expressed as:
\begin{equation}
\label{eqn:N}
    \overline{N}_w\approx 1+cd,\:\:d\in\mathbb{R}_0^+
\end{equation}

\noindent where $c$ is an interaction-density parameter with dimensions of inverse length. 
\end{hypothesis}

The coefficient $c$ is expected to be governed primarily by the beam and cavity geometry, although material-dependent cavity evolution may introduce state-wise and cross-material variations. Equation~\eqref{eqn:N} is intended as a leading-order depth representation of $\overline{N}_w$, rather than a point-wise approximation that requires $N_{\mathrm{eff}}=\overline{N}_w$. It is worth noting that $N_\mathrm{eff}$ remains governed by the complete ray-count distribution and the material-specific EPROD through Eq.~\eqref{eqn:cumulant_N}. 
\begin{hypothesis}
\label{hyp:exp}
Given Hypothesis~\ref{hyp:linear}, substituting the leading mean-count representation in Eq.~\eqref{eqn:N} into $N$ of Eq.~\eqref{eqn:eta_2} yields the following reduced-order approximate formulation of $\eta$ given $d$:
\begin{equation}
\label{eqn:eta_3}
    \eta\approx b\left[1-\exp{\left(-\tau_r\overline{N}_w\right)}\right]=b\left[1-\exp{\left(-\tau_r\right)}\exp{\left(-\tau_r cd\right)}\right]
\end{equation}

\noindent where $b$ serves as a beam-level closure factor that collectively accounts for the remaining difference between the leading mean-count representation and the complete optical response. 
\end{hypothesis}

Technically, $b$ includes the distributional separation between $\overline{N}_w$ and $N_{\mathrm{eff}}$, as well as incidence-angle-, temperature-, and surface-state-dependent optical effects not explicitly represented by the constant-EPROD approximation. The accuracy of Eq.~\eqref{eqn:eta_3} is therefore evaluated directly against the ray-resolved absorptance and experimental measurements.

For the convenience of analysis, we use $\mu\mathrm{m}^{-1}$ as the unit of $c$ in the subsequent text. The unit intercept in Eq.~\eqref{eqn:N} accounts for the initial laser--material interaction: even in the absence of vapor depression and subsequent internal reflections, the incident beam undergoes at least one interaction upon first impinging on the material surface. 

The following sections evaluate these relationships across simulation and experimental datasets, characterize the physical scales represented by $c$ and $b$, and examine their transfer behavior across material--beam conditions.

\section{Datasets and Methods}
\label{sec:methods}
\subsection{Ray-tracing simulation modeling and dataset}
\label{subsec:simulation}
\begin{figure*}[!ht]
\centering
\includegraphics[width=\linewidth]{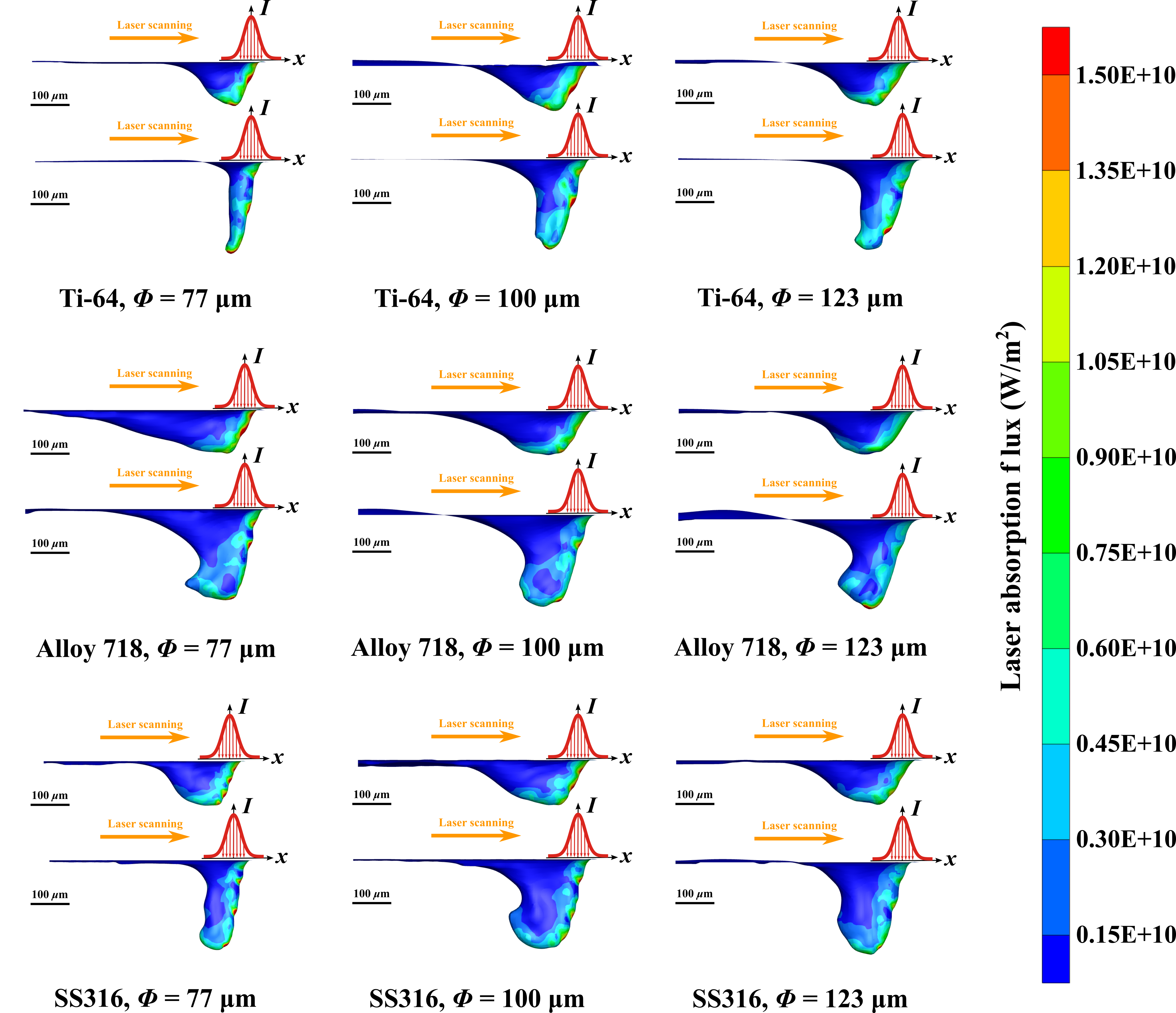}
\caption{Representative cavity geometries from the nine ray-resolved material--beam simulation groups. Rows correspond to Ti-64, Alloy~718, and SS316, and columns correspond to nominal spot diameters $\phi=77$, $100$, and $123$~$\mu$m. Each panel shows a shallower and a deeper cavity state together with the laser location, Gaussian intensity distribution, and scanning direction. Color denotes the local laser absorption flux (W$\cdot$m$^{-2}$). The variation among cavity states illustrates the morphology represented at leading order by depth and retained statistically through the ray-count distribution. }
\label{fig:simulation_geometry_3x3}
\end{figure*}

To interpret the ray reflection behavior and its relationship with keyhole geometry, ray-tracing simulations were performed using an existing multiphysics model developed for laser welding and additive manufacturing~\cite{kouraytem2019effect, tan2013investigation, tan2014analysis, li2021quantitative}. The model solves the conservation equations of mass, momentum, and energy to describe the thermo-fluid behavior of the melt-pool, while the evolving keyhole surface is captured using a level-set method. A ray-tracing module is coupled with the free-surface calculation to track the propagation, multiple reflections, and absorption of the incident laser rays on the keyhole surface. Keyhole morphologies with respect to varying conditions and materials are shown in Fig.~\ref{fig:simulation_geometry_3x3}. 

The incident laser beam is discretized into $640{,}000$ uniformly launched rays. As shown in Fig.~\ref{fig:simulation_geometry_3x3}, their powers follow a Gaussian radial profile, and the trajectory, interaction count, remaining power, and Fresnel absorption of each ray are tracked at the evolving liquid--vapor interface~\cite{kouraytem2019effect}. The dataset contains nine material--beam groups: Ti-64, Alloy~718, and SS316 at nominal spot diameters $\phi=77$, $100$, and $123~\mu\mathrm{m}$. For every sampled cavity state, we extracted the cavity depth $d$, incident-power-weighted mean interaction count $\overline{N}_w$, weighted variance $\sigma_{N,w}^{2}$, absorptance-equivalent count $N_{\mathrm{eff}}$, and total ray-traced absorptance $\eta$.

The archived effective single-interaction absorptances are $\varepsilon=0.385$ for Ti-64, $0.316$ for Alloy~718, and $0.341$ for SS316, corresponding to $0.4861$, $0.3798$, and $0.4170$ for EPROD, respectively. These values are prescribed inputs to the reduced-order analysis rather than fitted parameters~\cite{polyanskiy2024refractiveindex}. For each cavity state, $N_{\mathrm{eff}}$ is calculated from the complete power-weighted ray-count distribution using Eq.~\eqref{eqn:Neff_continuous} and the prescribed material-specific EPROD. It therefore represents the exact finite-beam aggregation under the constant-EPROD reference representation, whereas the full ray-tracing calculation retains the local incidence-angle (through its Fresnel treatment) and temperature dependence.

The reduced-order parameters were identified in two stages. First, $c$ was obtained from a constrained least-squares fit of Eq.~\eqref{eqn:N}, with the intercept fixed at one. Second, this fitted $c$ and the prescribed EPROD were held fixed, and only $b$ was fitted to the total-absorptance data using Eq.~\eqref{eqn:eta_3}. This sequence prevents the absorptance fit from trading changes in $b$ against changes in $c$ and preserves the interpretation of $c$ as the interaction-density coefficient inferred independently from the $\overline{N}_w$ statistics. The $N_{\mathrm{eff}}$ values are presented as a distribution-sensitive reference but are not used to identify $c$, because their explicit dependence on EPROD would combine geometric interaction accumulation with the prescribed material optical response. The coefficient of determination ($R^2$) and mean absolute percentage error (MAPE) were calculated for both fitting stages.

The multiphysics framework has previously been compared with melt-pool, cavity, reflected-radiation, and powder--gas observations under subsets of the materials and processing conditions considered here~\cite{deisenroth2026illuminating,kouraytem2019effect,li2021quantitative,li2020revealing,li2025effects,li2026elucidating}. These complementary comparisons provide physical grounding for using the simulated cavity states to resolve ray statistics and evaluate the proposed reduced-order relationships.

\subsection{Experimental dataset with keyhole imaging and laser absorptance measurements}
\label{subsec:exp}
The experimental evaluation uses synchronized Ti-64 absorptance and cavity-depth measurements obtained using a Gaussian laser spot diameter of $122.5\pm3~\mu\mathrm{m}$~\cite{SIMONDS2020775,simonds2021causal}. Total absorptance was measured using integrating-sphere radiometry while the evolving cavity geometry was recorded using high-speed synchrotron X-ray imaging. These paired measurements provide a direct experimental evaluation of the composite absorptance--depth relation in Eq.~\eqref{eqn:eta_3}. Because individual ray histories, $\overline{N}_w$, and $N_{\mathrm{eff}}$ are not available experimentally, the simulation-stage identification of $c$ from the interaction-count relation cannot be applied directly to this dataset. Instead, $c$ and $b$ were fitted simultaneously to the measured $(d,\eta)$ pairs using Eq.~\eqref{eqn:eta_3}, with $\varepsilon=0.385$ and $\tau_r=0.4861$ prescribed for Ti-64. Neither fitted parameter was constrained by the simulation. The resulting experimental $c$ is therefore an effective coefficient of the composite absorptance--depth relation rather than a direct measurement of either interaction-count statistic. Its comparison with the independently determined simulation coefficient evaluates whether a consistent characteristic depth scale emerges through the two identification routes.

\section{Results and discussion}
\label{sec:results}
\subsection{Ray-resolved simulation analysis}
\label{subsec:simulation_results}
\subsubsection{Interaction-count statistics versus cavity depth}
\begin{figure*}[!ht]
    \centering
    \begin{subfigure}[b]{0.32\textwidth}
        \centering
        \includegraphics[width=\textwidth]{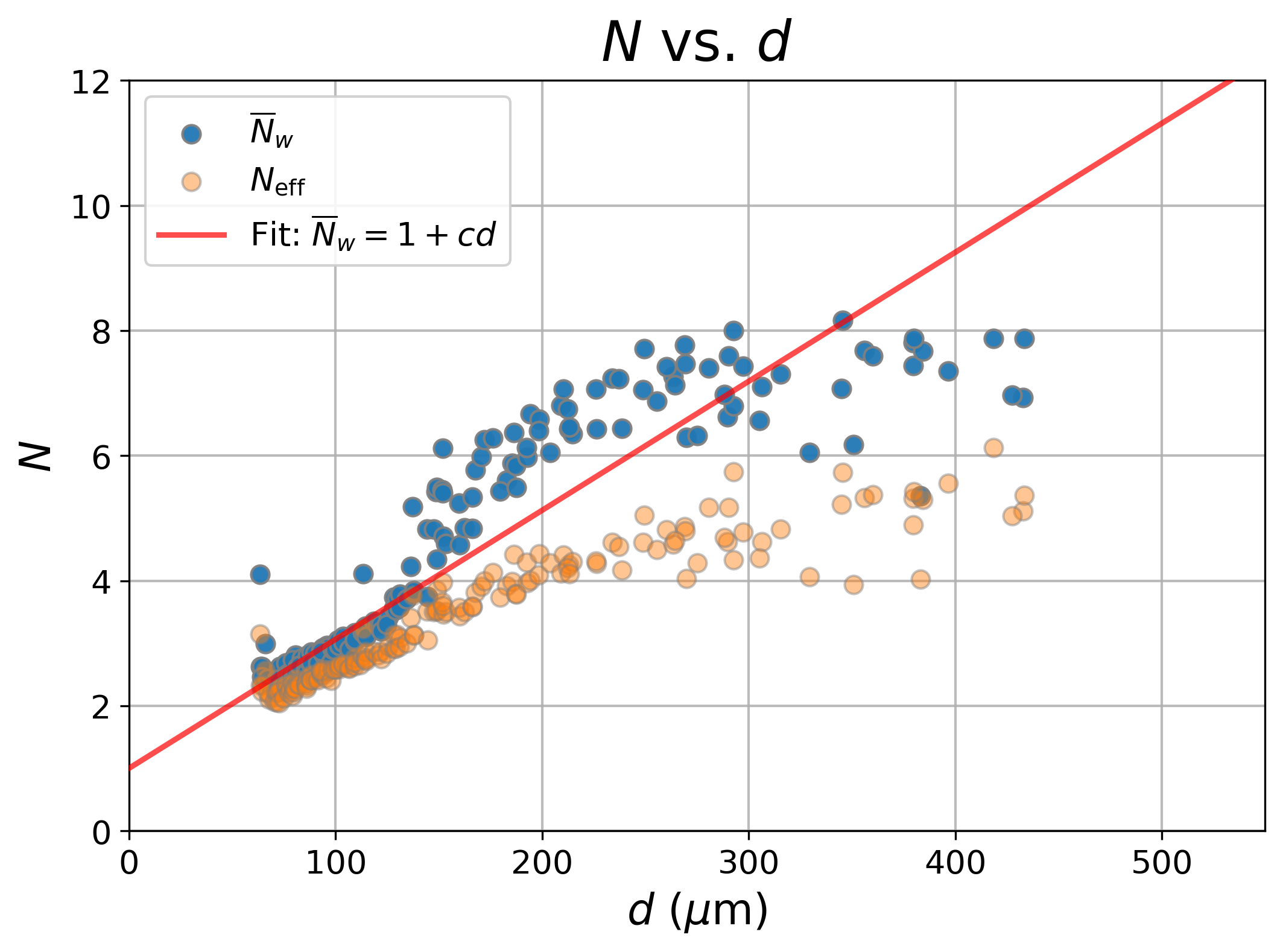}
        \caption{Ti-64, $\phi=77~\mu\mathrm{m}$}
        \label{subfig:sim_Nvsd_Ti64-77}
    \end{subfigure}
    \hfill
    \begin{subfigure}[b]{0.32\textwidth}
        \centering
        \includegraphics[width=\textwidth]{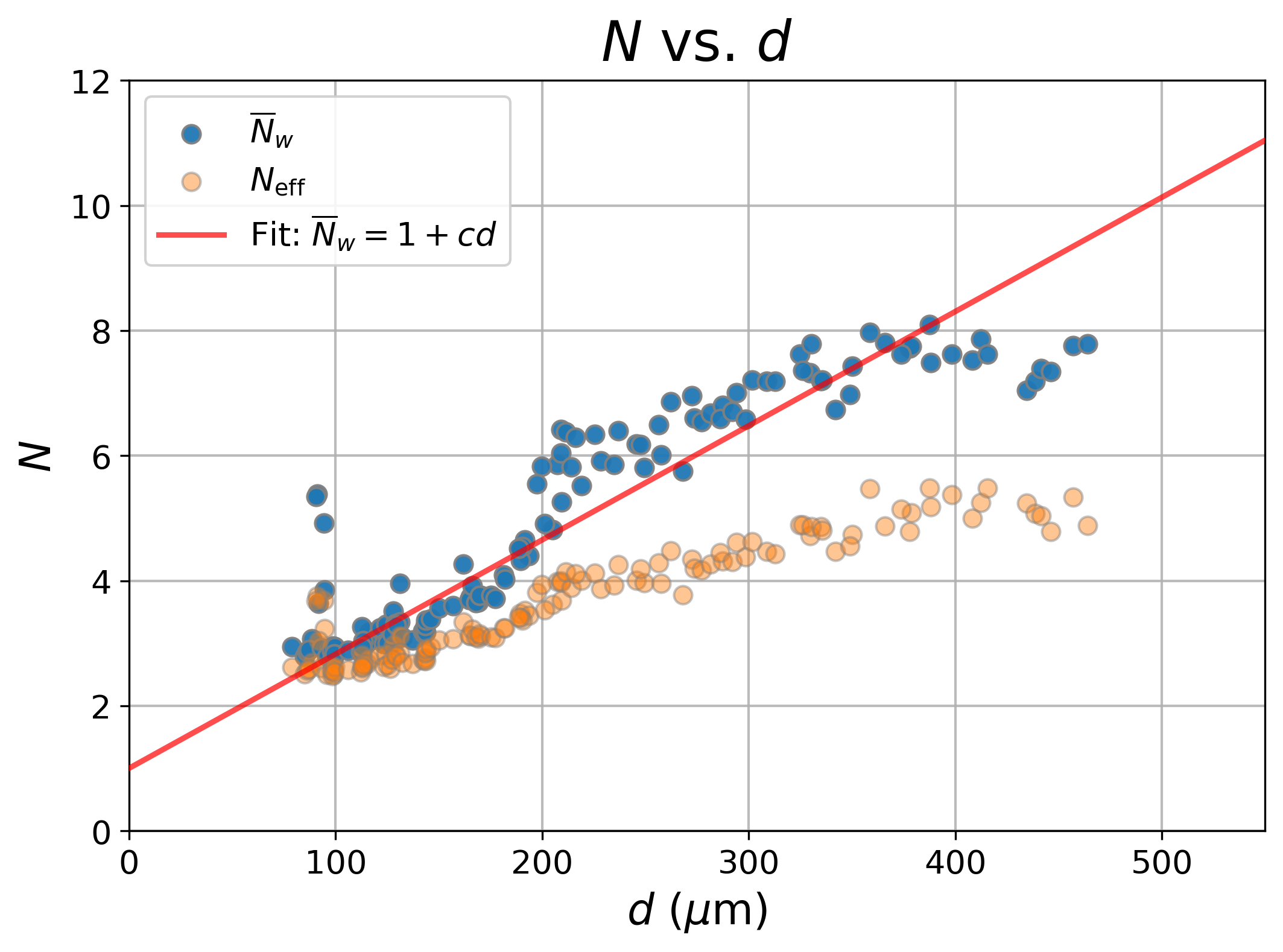}
        \caption{Ti-64, $\phi=100~\mu\mathrm{m}$}
        \label{fig:sim_Nvsd_Ti64-100}
    \end{subfigure}
    \hfill
    \begin{subfigure}[b]{0.32\textwidth}
        \centering
        \includegraphics[width=\textwidth]{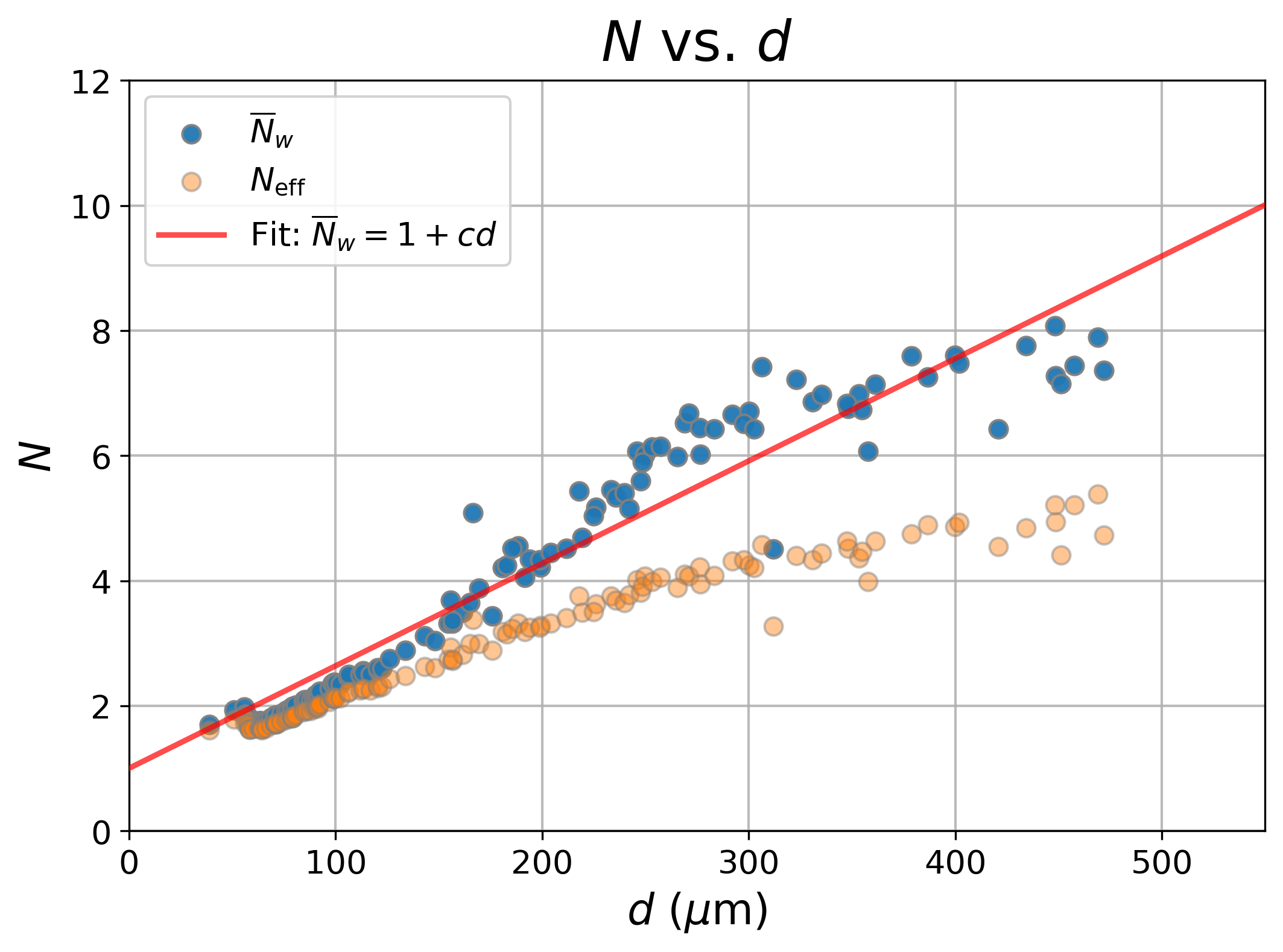}
        \caption{Ti-64, $\phi=123~\mu\mathrm{m}$}
        \label{fig:sim_Nvsd_Ti64-123}
    \end{subfigure}

    \vspace{0.3cm}
    
    \begin{subfigure}[b]{0.32\textwidth}
        \centering
        \includegraphics[width=\textwidth]{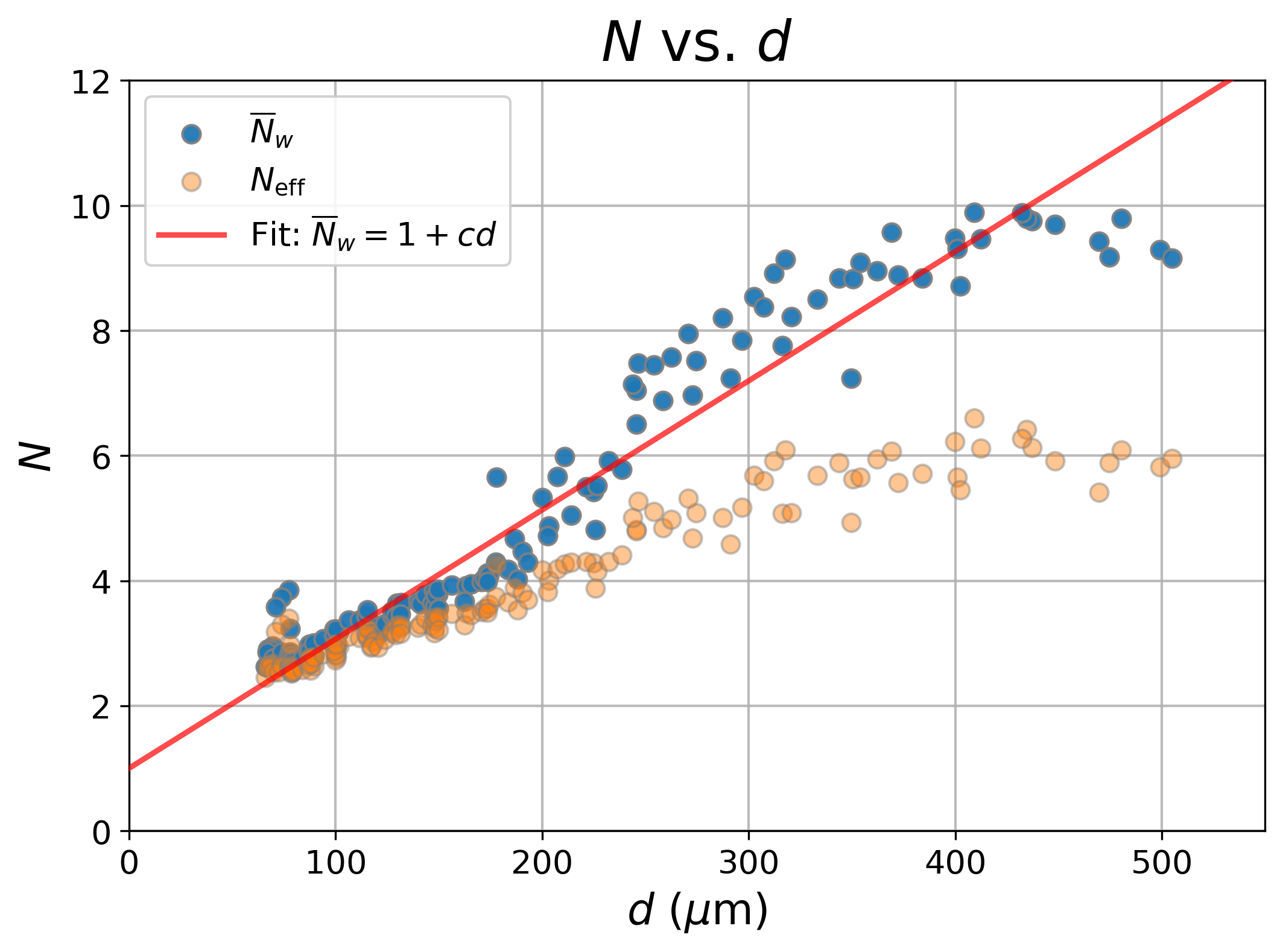}
        \caption{Alloy~718, $\phi=77~\mu\mathrm{m}$}
        \label{subfig:sim_Nvsd_IN-77}
    \end{subfigure}
    \hfill
    \begin{subfigure}[b]{0.32\textwidth}
        \centering
        \includegraphics[width=\textwidth]{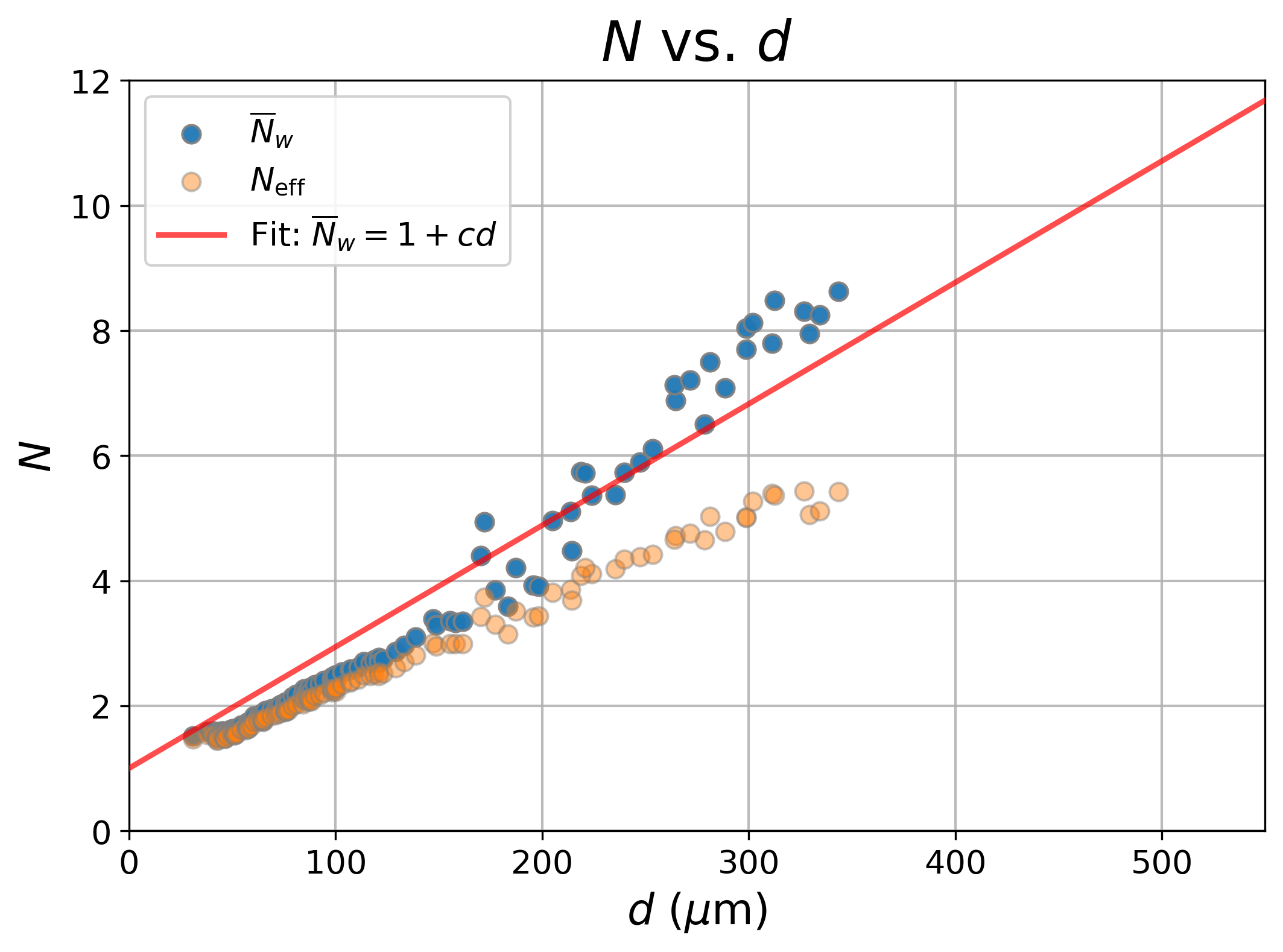}
        \caption{Alloy~718, $\phi=100~\mu\mathrm{m}$}
        \label{subfig:sim_Nvsd_IN-100}
    \end{subfigure}
    \hfill
    \begin{subfigure}[b]{0.32\textwidth}
        \centering
        \includegraphics[width=\textwidth]{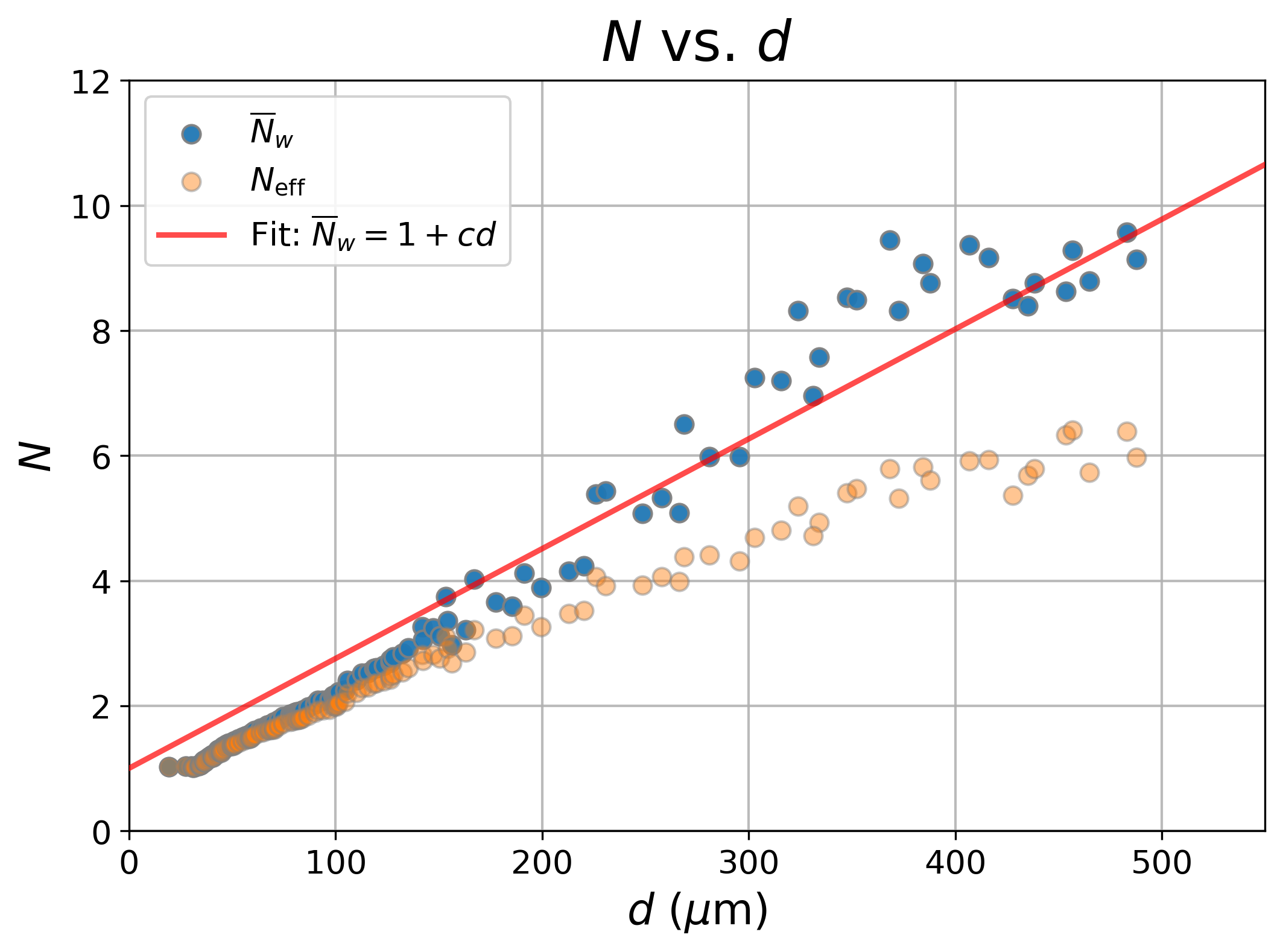}
        \caption{Alloy~718, $\phi=123~\mu\mathrm{m}$}
        \label{subfig:sim_Nvsd_IN-123}
    \end{subfigure}

    \vspace{0.3cm}

    \begin{subfigure}[b]{0.32\textwidth}
        \centering
        \includegraphics[width=\textwidth]{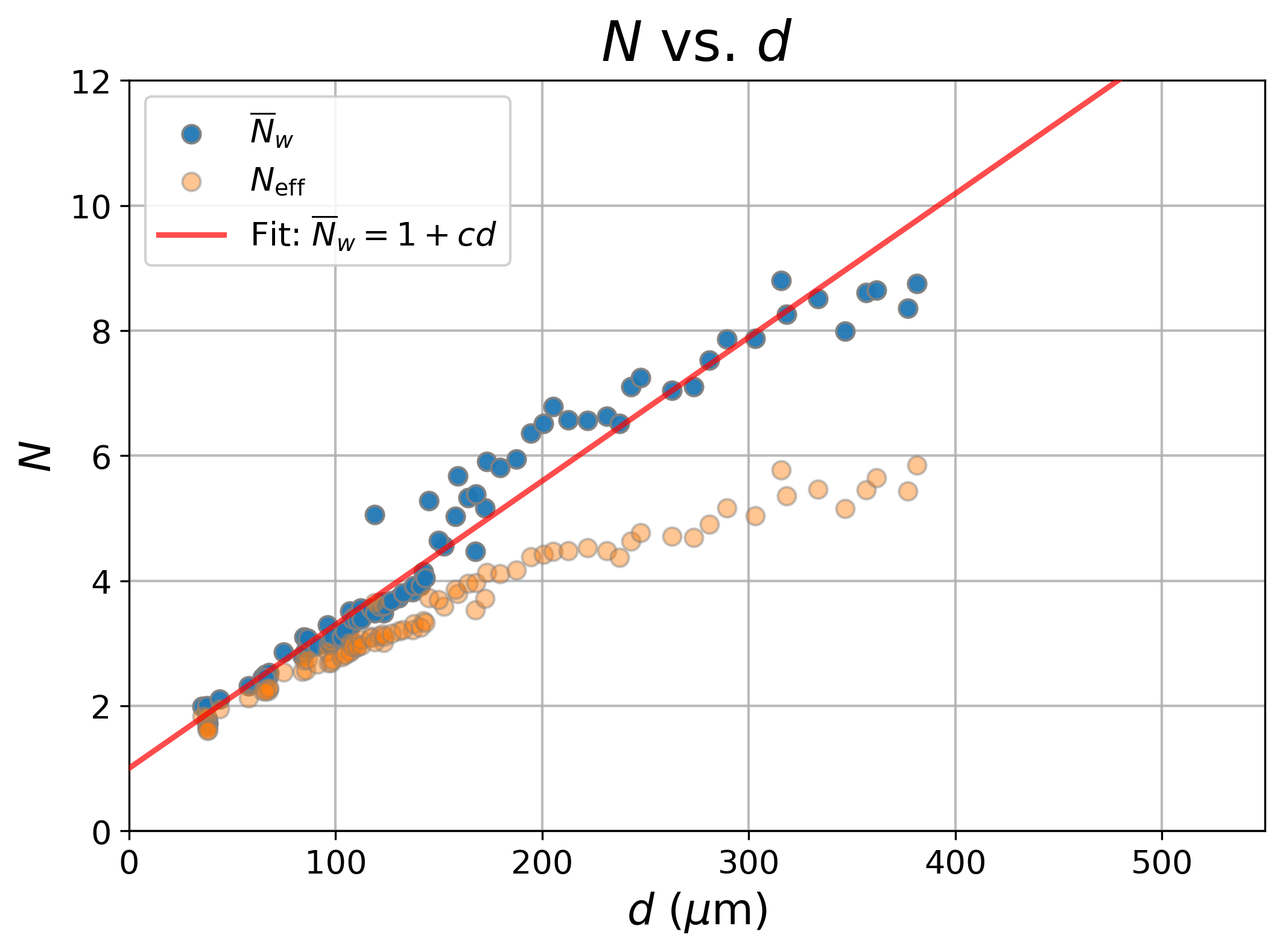}
        \caption{SS316, $\phi=77~\mu\mathrm{m}$}
        \label{subfig:sim_Nvsd_SS316-77}
    \end{subfigure}
    \hfill
    \begin{subfigure}[b]{0.32\textwidth}
        \centering
        \includegraphics[width=\textwidth]{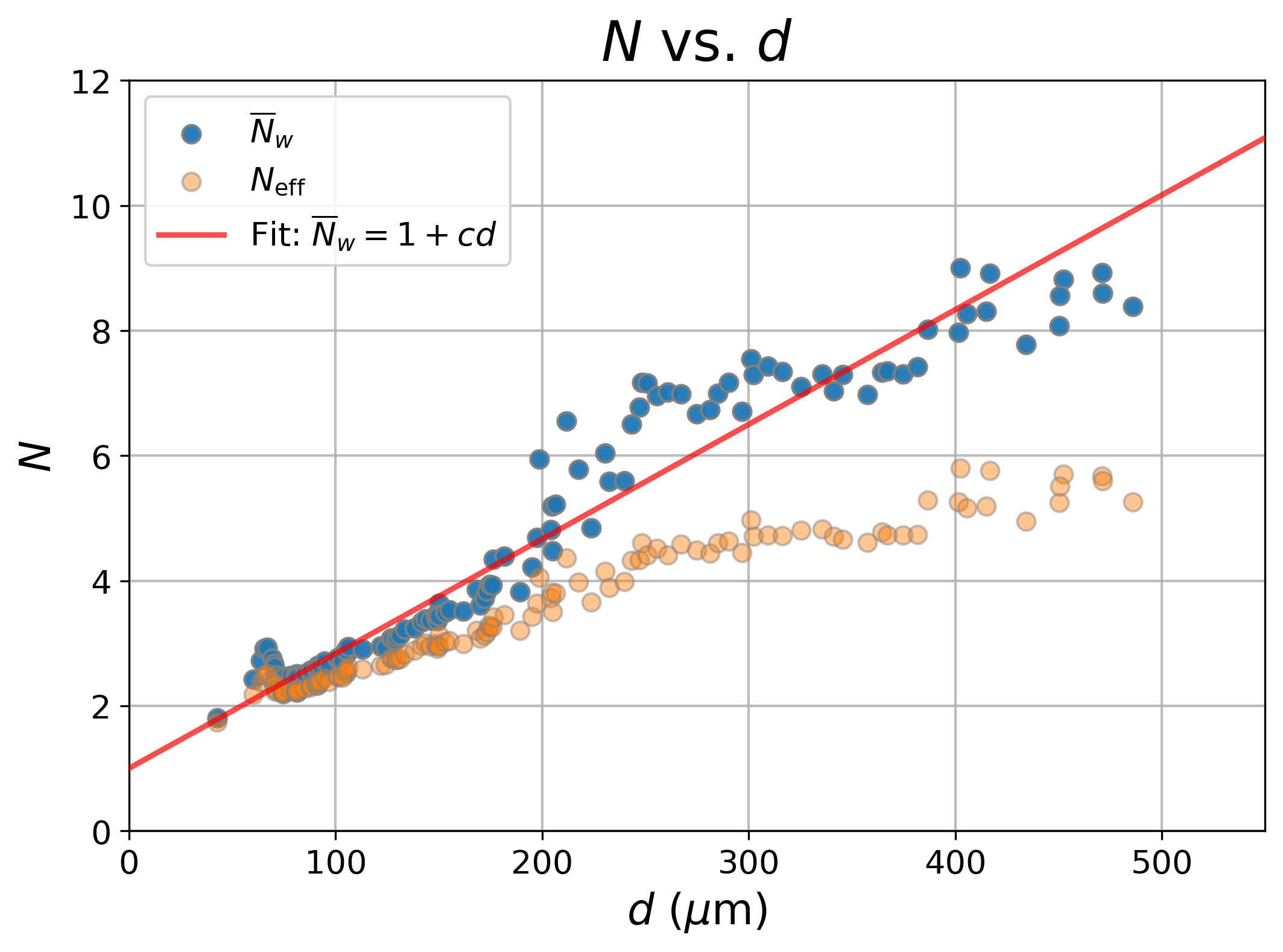}
        \caption{SS316, $\phi=100~\mu\mathrm{m}$}
        \label{subfig:sim_Nvsd_SS316-100}
    \end{subfigure}
    \hfill
    \begin{subfigure}[b]{0.32\textwidth}
        \centering
        \includegraphics[width=\textwidth]{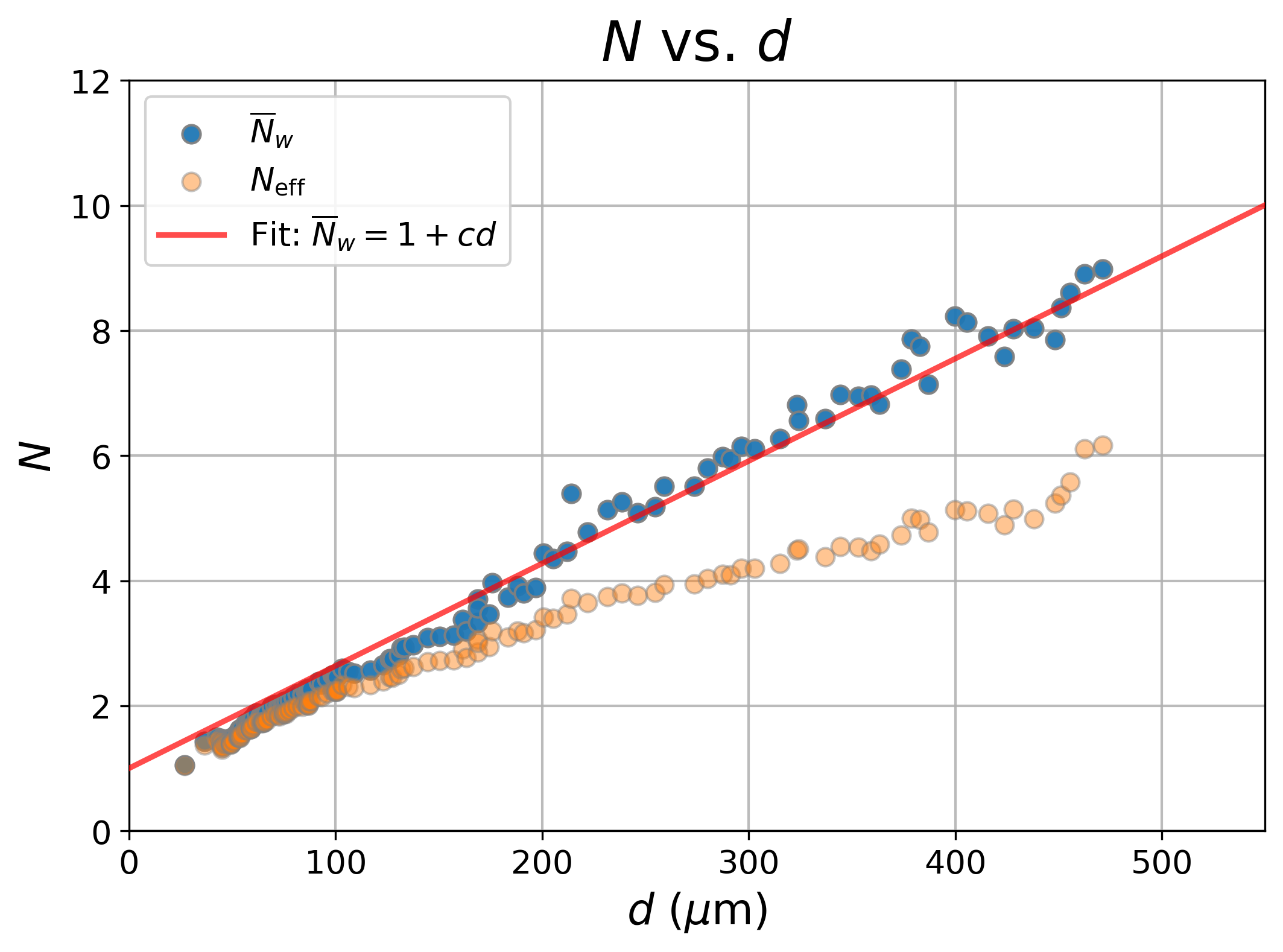}
        \caption{SS316, $\phi=123~\mu\mathrm{m}$}
        \label{subfig:sim_Nvsd_SS316-123}
    \end{subfigure}
    
    \caption{Power-weighted mean interaction count, $\overline{N}_w$, and absorptance-equivalent interaction count, $N_{\mathrm{eff}}$, as functions of cavity depth, $d$, for all nine ray-resolved simulation groups. Blue points denote $\overline{N}_w$, orange points denote $N_{\mathrm{eff}}$ calculated using the prescribed material-specific EPROD, and red lines show the constrained leading-order fits $\overline{N}_w=1+cd$. The $N_{\mathrm{eff}}$ values visualize the distribution- and EPROD-dependent correction to the mean interaction count and are not used to identify $c$. Rows show Ti-64, Alloy~718, and SS316; columns show $\phi=77$, $100$, and $123~\mu\mathrm{m}$.}
    \label{fig:sim_Nvsd}
\end{figure*}

Figure~\ref{fig:sim_Nvsd} evaluates Eq.~\eqref{eqn:N} while also showing the corresponding absorptance-equivalent interaction counts for all nine groups; Tab.~\ref{tab:sim_Nvsd} reports the fitted coefficients and errors for $\overline{N}_w$. The unit-intercept relation captures the dominant increase in $\overline{N}_w$ with $d$, with $R^2=0.7824$--$0.9843$ and MAPE$=6.39$--$27.92\%$. Seven of the nine conditions have $R^2\ge0.9312$. Because the simulated states do not densely sample $d=0$, the analysis primarily evaluates the slope $c$; the unit intercept remains a physically imposed boundary condition.

The $N_{\mathrm{eff}}$ values also increase smoothly with depth and appear approximately linear over substantial portions of the sampled ranges. This behavior is consistent with both statistics being generated by the same evolving ray-trajectory field, but it does not make their depth-scaling coefficients equivalent. The apparent slope and intercept of $N_{\mathrm{eff}}$ remain conditioned by EPROD and by the complete ray-count distribution, whereas $c$ is defined from $\overline{N}_w$. As required by Jensen's inequality, $N_{\mathrm{eff}}$ remains below $\overline{N}_w$ across the sampled states. In several conditions, particularly Ti-64 at a spot diameter of $77~\mu\mathrm{m}$, the distribution-sensitive correction produces a smoother $N_{\mathrm{eff}}$ trend than the corresponding arithmetic mean. This behavior is consistent with Eq.~\eqref{eqn:cumulant_N}: highly confined, large-count ray families can strongly increase $\overline{N}_w$, while their influence on the surviving beam power is exponentially suppressed in $N_{\mathrm{eff}}$. The comparison therefore illustrates the role of ray-path heterogeneity without changing the use of $\overline{N}_w$ as the EPROD-independent statistic for identifying $c$.

\begin{table}[!htbp]
\centering
\caption{Unit-intercept fits of $\overline{N}_w=1+cd$ for each simulated material and laser spot diameter $\phi$. The prescribed effective single-interaction absorptance $\varepsilon$ is listed with each material.}
\label{tab:sim_Nvsd}
\begingroup
\small
\renewcommand{\arraystretch}{1.20}
\setlength{\tabcolsep}{3.5pt}

\sisetup{
    detect-weight=true,
    detect-family=true,
    table-number-alignment=center,
    table-text-alignment=center
}

\begin{tabular*}{\linewidth}{
    @{\hspace{0.6em}\extracolsep{\fill}}
    l
    *{3}{
        S[table-format=1.4]
        S[table-format=1.4]
        S[table-format=2.2]
    }
    @{\hspace{0.6em}}
}
\toprule

\multicolumn{1}{c}{%
    \multirow[c]{3}{*}{%
        \makecell[c]{\bfseries\boldmath
        $d$~\textit{vs.}~$\overline{N}_w$\\[-0.2ex]simulation}}}
& \multicolumn{3}{c}{$\phi=77~\mu\mathrm{m}$}
& \multicolumn{3}{c}{$\phi=100~\mu\mathrm{m}$}
& \multicolumn{3}{c}{$\phi=123~\mu\mathrm{m}$} \\
\cmidrule(lr){2-4}
\cmidrule(lr){5-7}
\cmidrule(lr){8-10}

& \multicolumn{1}{c}{$c$}
& \multicolumn{1}{c}{\multirow[c]{2}{*}{$R^2$}}
& \multicolumn{1}{c}{MAPE}
& \multicolumn{1}{c}{$c$}
& \multicolumn{1}{c}{\multirow[c]{2}{*}{$R^2$}}
& \multicolumn{1}{c}{MAPE}
& \multicolumn{1}{c}{$c$}
& \multicolumn{1}{c}{\multirow[c]{2}{*}{$R^2$}}
& \multicolumn{1}{c}{MAPE} \\[-0.2ex]

& \multicolumn{1}{c}{$\left(\mu\mathrm{m}^{-1}\right)$}
&
& \multicolumn{1}{c}{(\%)}
& \multicolumn{1}{c}{$\left(\mu\mathrm{m}^{-1}\right)$}
&
& \multicolumn{1}{c}{(\%)}
& \multicolumn{1}{c}{$\left(\mu\mathrm{m}^{-1}\right)$}
&
& \multicolumn{1}{c}{(\%)} \\

\midrule

\makecell[l]{Ti-64\\[-0.2ex]($\varepsilon=0.385$)}
& 0.0206
& 0.7824
& 11.05
& 0.0183
& 0.8293
& 9.77
& 0.0164
& 0.9358
& 11.84 \\
\addlinespace[2pt]

\makecell[l]{Alloy~718\\[-0.2ex]($\varepsilon=0.316$)}
& 0.0207
& 0.9318
& 8.62
& 0.0194
& 0.9312
& 17.82
& 0.0176
& 0.9467
& 27.92 \\
\addlinespace[2pt]

\makecell[l]{SS316\\[-0.2ex]($\varepsilon=0.341$)}
& 0.0230
& 0.9522
& 6.39
& 0.0183
& 0.9381
& 7.57
& 0.0164
& 0.9843
& 10.14 \\

\bottomrule
\end{tabular*}
\endgroup
\end{table}

\begin{figure}[!ht]
\centering
\includegraphics[width=0.8\linewidth]{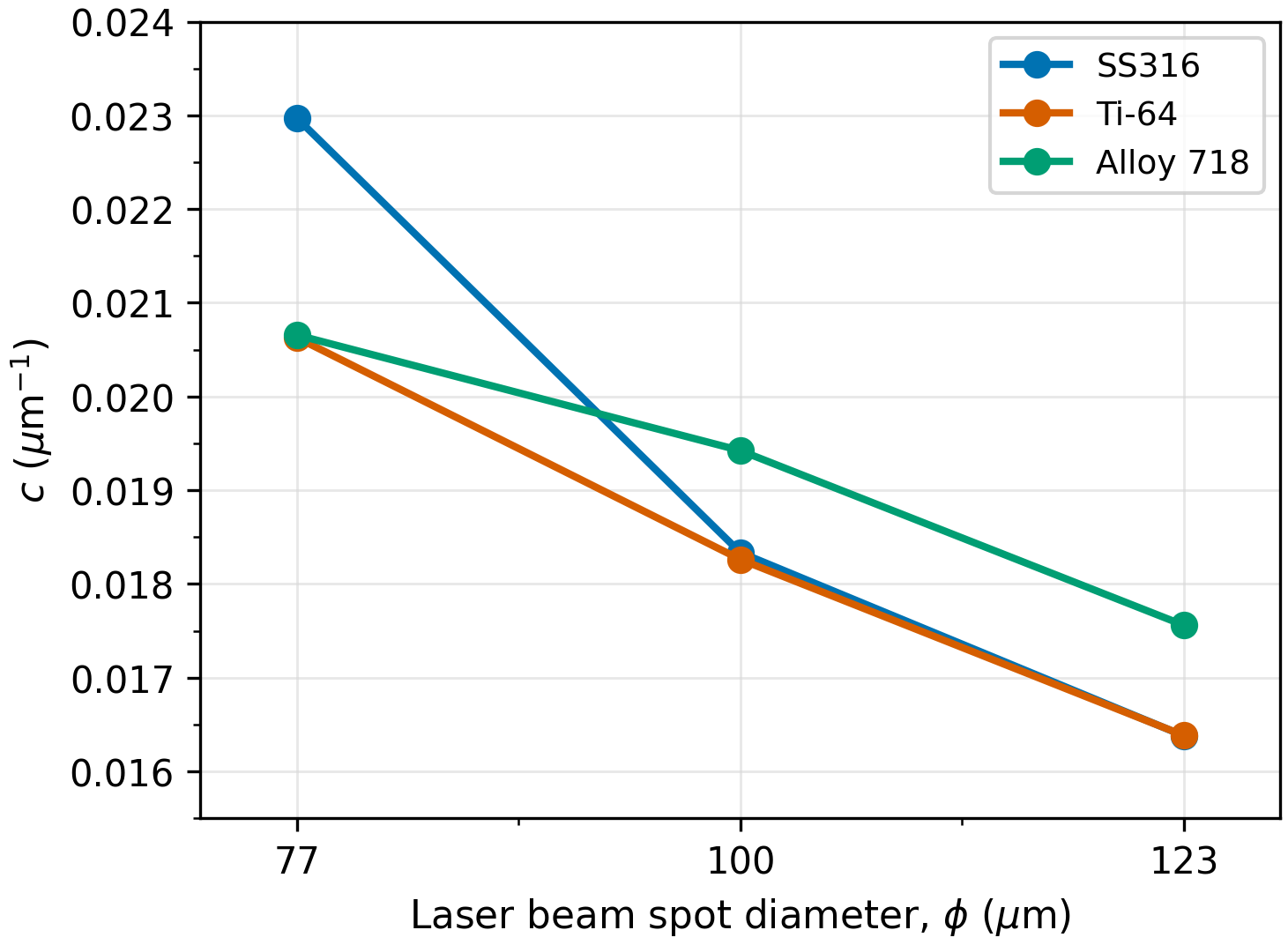}
\caption{Dependence of the fitted depth-scaling coefficient $c$ on the laser-beam spot diameter $\phi$ for Ti-64, Alloy~718, and SS316. For each material and spot diameter, $c$ is obtained by fitting the unit-intercept relation $\overline{N}_w=1+cd$ to the ray-resolved simulation data. Across the sampled range, $c$ decreases monotonically with increasing $\phi$ for all three materials, corresponding to a weaker increase in $\overline{N}_w$ per unit keyhole depth for wider laser beams. }
\label{fig:c_vs_phi}
\end{figure}

The fitted coefficients span $c=0.0164$--$0.0230~\mu\mathrm{m}^{-1}$, corresponding to one additional power-weighted mean interaction per approximately $43$--$61~\mu\mathrm{m}$ of cavity depth. At fixed spot diameter, the cross-alloy coefficient of variation is only $3.4$--$6.3\%$. By comparison, increasing $\phi$ from $77$ to $123~\mu\mathrm{m}$ reduces $c$ by approximately $20\%$ for Ti-64, $15\%$ for Alloy~718, and $29\%$ for SS316, as shown in Fig.~\ref{fig:c_vs_phi}. Beam size therefore produces the stronger systematic variation within the present matrix. Because $\overline{N}_w$ contains no explicit EPROD dependence, these results support interpreting $c$ as an approximately alloy-stable, beam- and morphology-conditioned geometric interaction scale, while EPROD carries the explicit material-optical contribution.

The distinction between $\overline{N}_w$ and $N_{\mathrm{eff}}$ arises from heterogeneity in the power-weighted distribution of ray interaction counts, which is governed by the evolving cavity morphology illustrated in Fig.~\ref{fig:simulation_geometry_3x3}. Meanwhile, Fig.~\ref{fig:sim_Nvsd} reveals condition-dependent depth ranges in which the simulated $\overline{N}_w$ departs from the leading-order relation $\overline{N}_w=1+cd$, consistent with observations reported in previous work~\cite{allen2020energy}. These localized departures indicate that cavity depth captures the dominant accumulation of laser--wall interactions but does not uniquely represent changes in opening geometry, wall inclination, curvature, and the resulting organization of ray trajectories. The simultaneous presentation of $\overline{N}_w$ and $N_{\mathrm{eff}}$ shows how these distributional changes can appear strongly in the arithmetic mean while being partially suppressed in the absorptance-equivalent statistic. Overall, the unit-intercept relation provides a consistent and physically interpretable first-order description of the EPROD-independent interaction scale across the material--beam matrix.

\subsubsection{Ray-count heterogeneity and the effective-count correction}
After testing the $d$~\textit{vs.}~$\overline{N}_w$ hypothesis, we quantified the leading difference between the mean interaction count and the absorptance-equivalent count. Normalizing the second-order term in Eq.~\eqref{eqn:cumulant_N} by $\overline{N}_w$ gives:
\begin{equation}
\label{eqn:relative_variance_correction}
\sigma_N^{(2)}(d)=\frac{\tau_r\sigma_{N,w}^{2}}{2\overline{N}_w}
\end{equation}

\begin{figure*}[!htbp]
    \centering
    \begin{subfigure}[b]{0.32\textwidth}
        \centering
        \includegraphics[width=\textwidth]{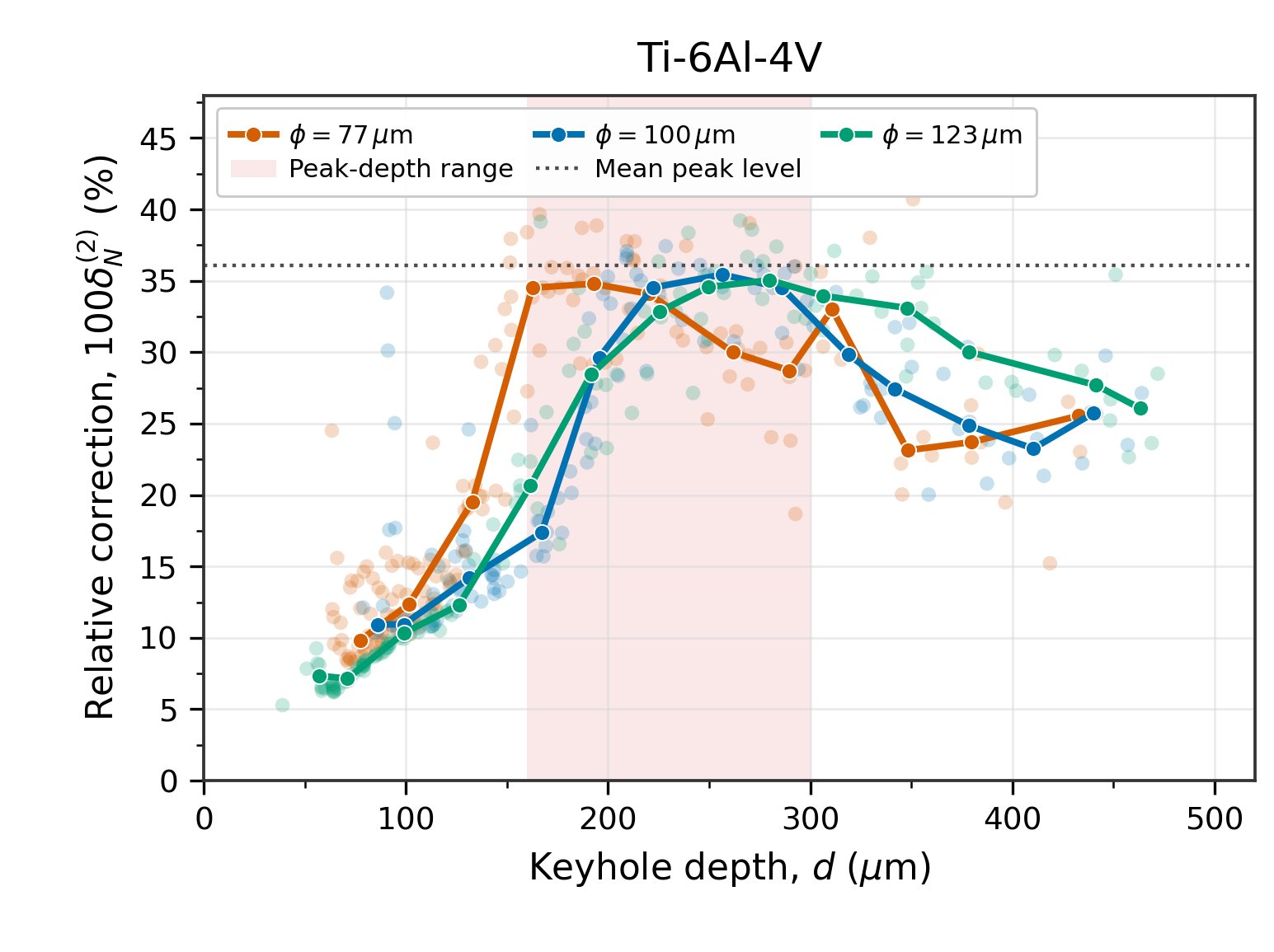}
        \caption{Ti-64}
        \label{subfig:sim_variance_Ti64}
    \end{subfigure}
    \hfill
    \begin{subfigure}[b]{0.32\textwidth}
        \centering
        \includegraphics[width=\textwidth]{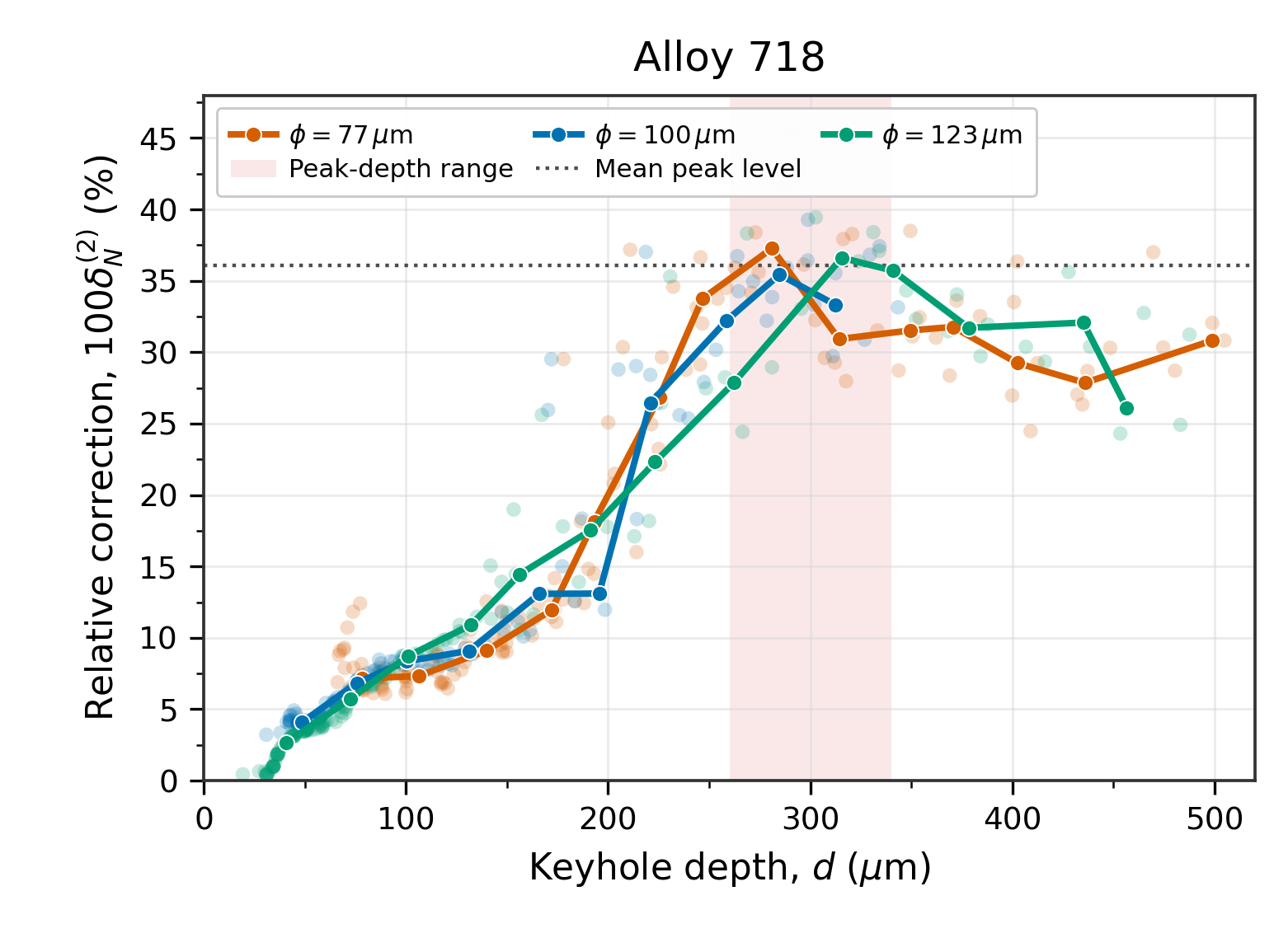}
        \caption{Alloy~718}
        \label{fig:sim_variance_IN}
    \end{subfigure}
    \hfill
    \begin{subfigure}[b]{0.32\textwidth}
        \centering
        \includegraphics[width=\textwidth]{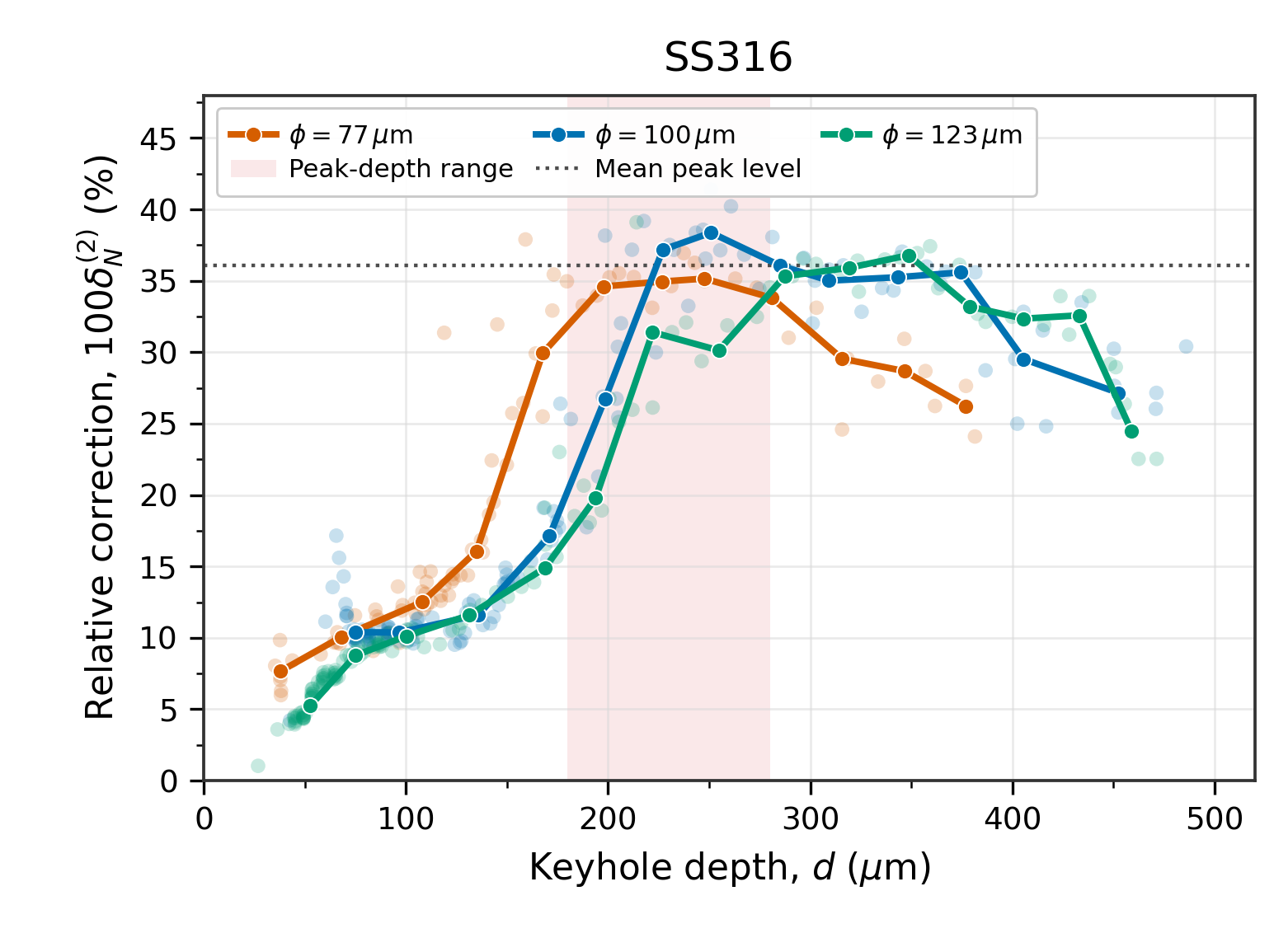}
        \caption{SS316}
        \label{fig:sim_variance_SS}
    \end{subfigure}
    \caption{Relative second-order correction $\sigma_N^{(2)}$ as a function of cavity depth for Ti-64, Alloy~718, and SS316. Within each panel, orange, blue, and green denote spot diameters $\phi=77$, $100$, and $123~\mu\mathrm{m}$, respectively. Translucent points represent individual simulated cavity states, and the opaque line-and-marker traces summarize the depth-dependent trends. Shaded regions indicate the reported ranges containing the condition-wise maxima, and the dotted horizontal line denotes their mean peak level. These graphical guides are descriptive and do not define a universal transition threshold.}
    \label{fig:variance_correction}
\end{figure*}

This dimensionless quantity measures the fractional reduction from $\overline{N}_w$ to $N_{\mathrm{eff}}$ contributed by the second-order variance term. Across all nine material--spot-diameter conditions, Fig.~\ref{fig:variance_correction} reveals a consistent depth-dependent evolution. In the shallow-cavity portion of the present simulations, approximately $d\lesssim100~\mu\mathrm{m}$, the correction is generally about $15\%$ or less, providing an empirical range in which $N_{\mathrm{eff}}\approx\overline{N}_w$ is a useful leading approximation. The correction increases markedly as the cavity develops, reaches a broad maximum or plateau over a condition-dependent intermediate-depth range of roughly $150$--$350~\mu\mathrm{m}$, and generally decreases at greater depths. Although the depth at which the maximum occurs and the ordering among spot diameters vary across conditions, the peak magnitudes cluster within a relatively narrow range of approximately $33\%$--$38\%$, with a cross-condition mean near $36\%$. This depth criterion concerns the approximation $N_{\mathrm{eff}}\approx\overline{N}_w$, rather than the broader leading-order relation $\overline{N}_w=1+cd$, which we evaluate over the full sampled depth range. The red-shaded region in each panel denotes the material-specific envelope spanning the peak-depth locations of the three spot-diameter curves.

These peak-depth ranges broadly coincide with the intervals in which the corresponding $\overline{N}_w$--$d$ data exhibit their most pronounced localized upward departures from the leading-order relation $\overline{N}_w=1+cd$ in Fig.~\ref{fig:sim_Nvsd}. This co-occurrence suggests that the enhanced ray-count heterogeneity and the departure of the mean interaction count from its leading-order depth dependence are complementary statistical signatures of a geometry-driven reorganization of ray trajectories. Such a reorganization may involve the simultaneous presence of relatively short escape paths and more strongly confined trajectories that undergo additional wall interactions. Because the second-order correction lowers $N_{\mathrm{eff}}$ relative to $\overline{N}_w$, it can partially counterbalance localized upward departures of the mean count from its fitted linear trend. This explains why $N_{\mathrm{eff}}$ may exhibit a smoother depth dependence even when $\overline{N}_w$ shows pronounced statewise excursions. The two statistics are therefore not numerically interchangeable: $\overline{N}_w$ supplies the EPROD-independent leading geometric scale used to define $c$, whereas $N_{\mathrm{eff}}$ quantifies how material-conditioned optical aggregation modifies that scale. Beyond the peak region, the normalized second-order correction generally weakens, bringing the two quantities fractionally closer within the cumulant description.

\subsubsection{Total absorptance versus cavity depth}
\begin{figure*}[!ht]
    \centering
    \begin{subfigure}[b]{0.32\textwidth}
        \centering
        \includegraphics[width=\textwidth]{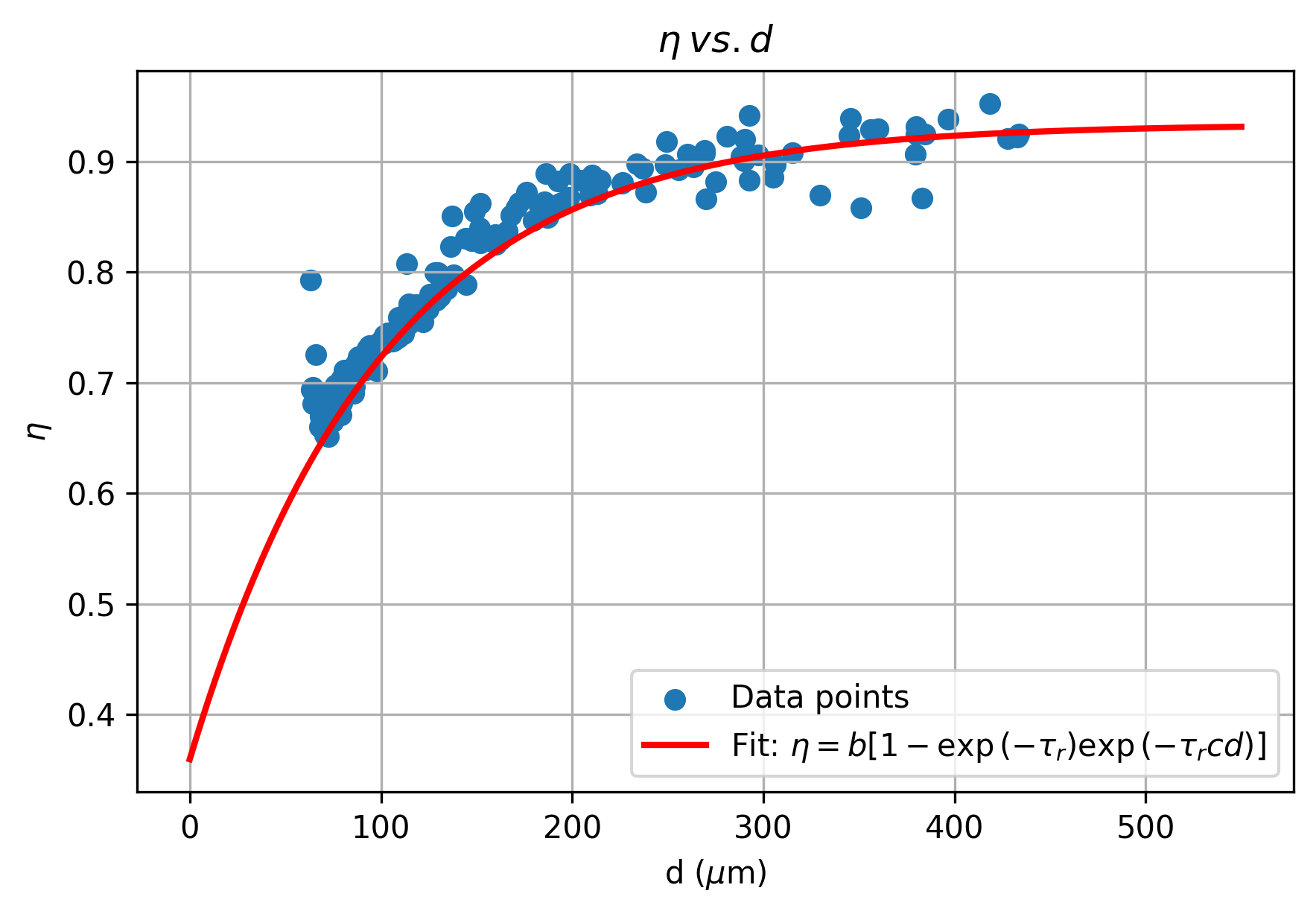}
        \caption{Ti-64, $\phi=77~\mu\mathrm{m}$}
        \label{subfig:sim_etavsd_Ti64-77}
    \end{subfigure}
    \hfill
    \begin{subfigure}[b]{0.32\textwidth}
        \centering
        \includegraphics[width=\textwidth]{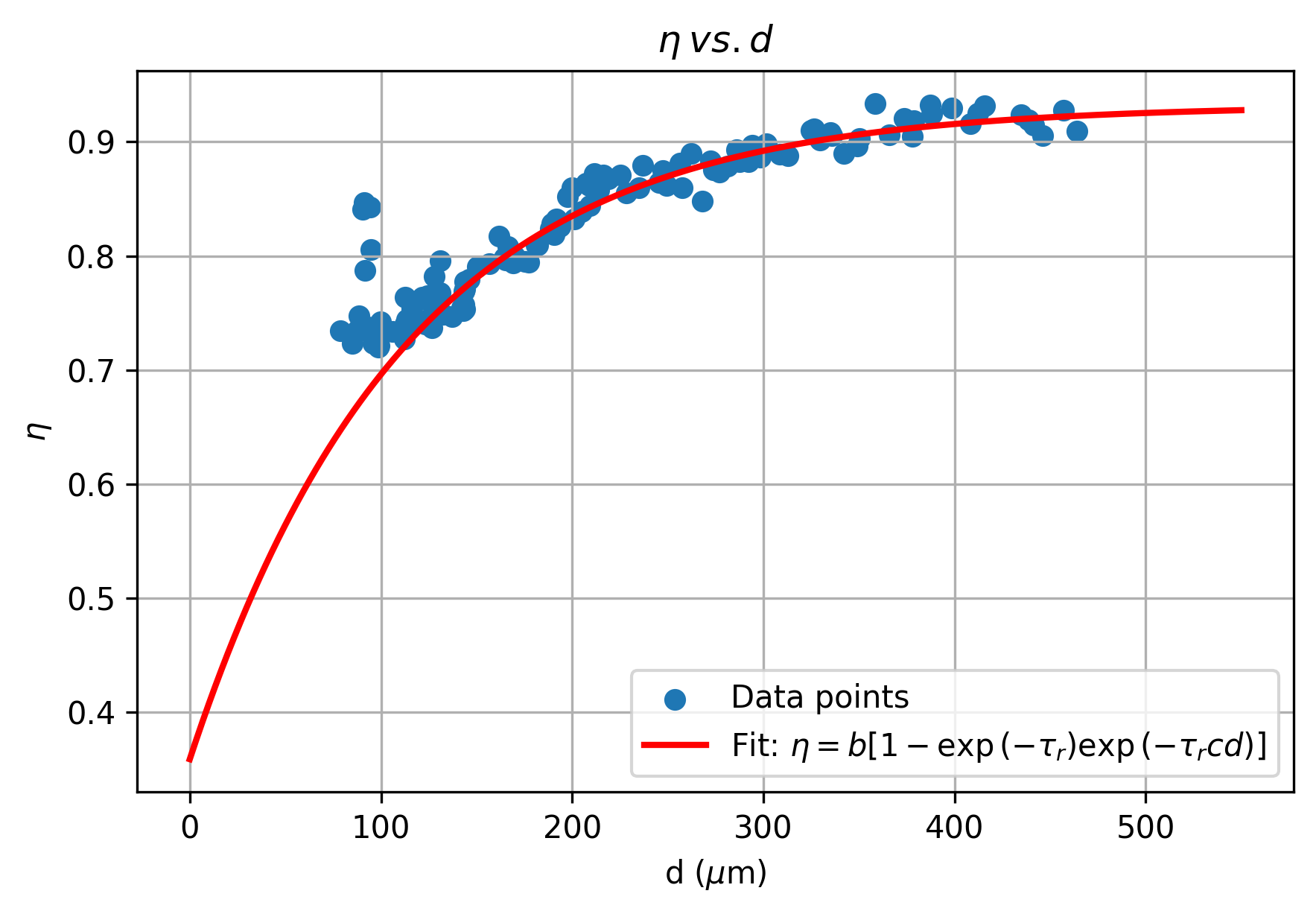}
        \caption{Ti-64, $\phi=100~\mu\mathrm{m}$}
        \label{fig:sim_etavsd_Ti64-100}
    \end{subfigure}
    \hfill
    \begin{subfigure}[b]{0.32\textwidth}
        \centering
        \includegraphics[width=\textwidth]{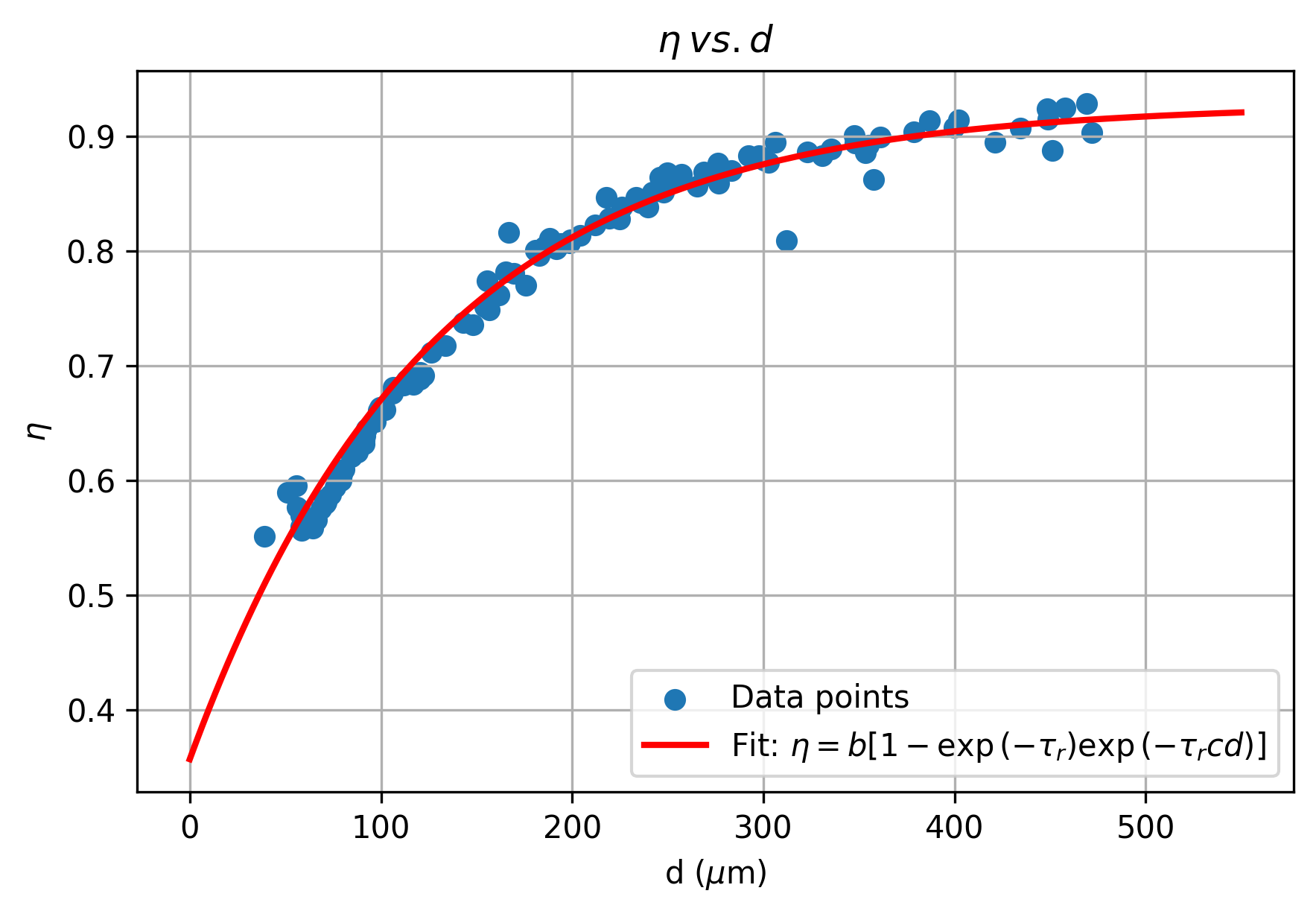}
        \caption{Ti-64, $\phi=123~\mu\mathrm{m}$}
        \label{fig:sim_etavsd_Ti64-123}
    \end{subfigure}

    \vspace{0.3cm}
    
    \begin{subfigure}[b]{0.32\textwidth}
        \centering
        \includegraphics[width=\textwidth]{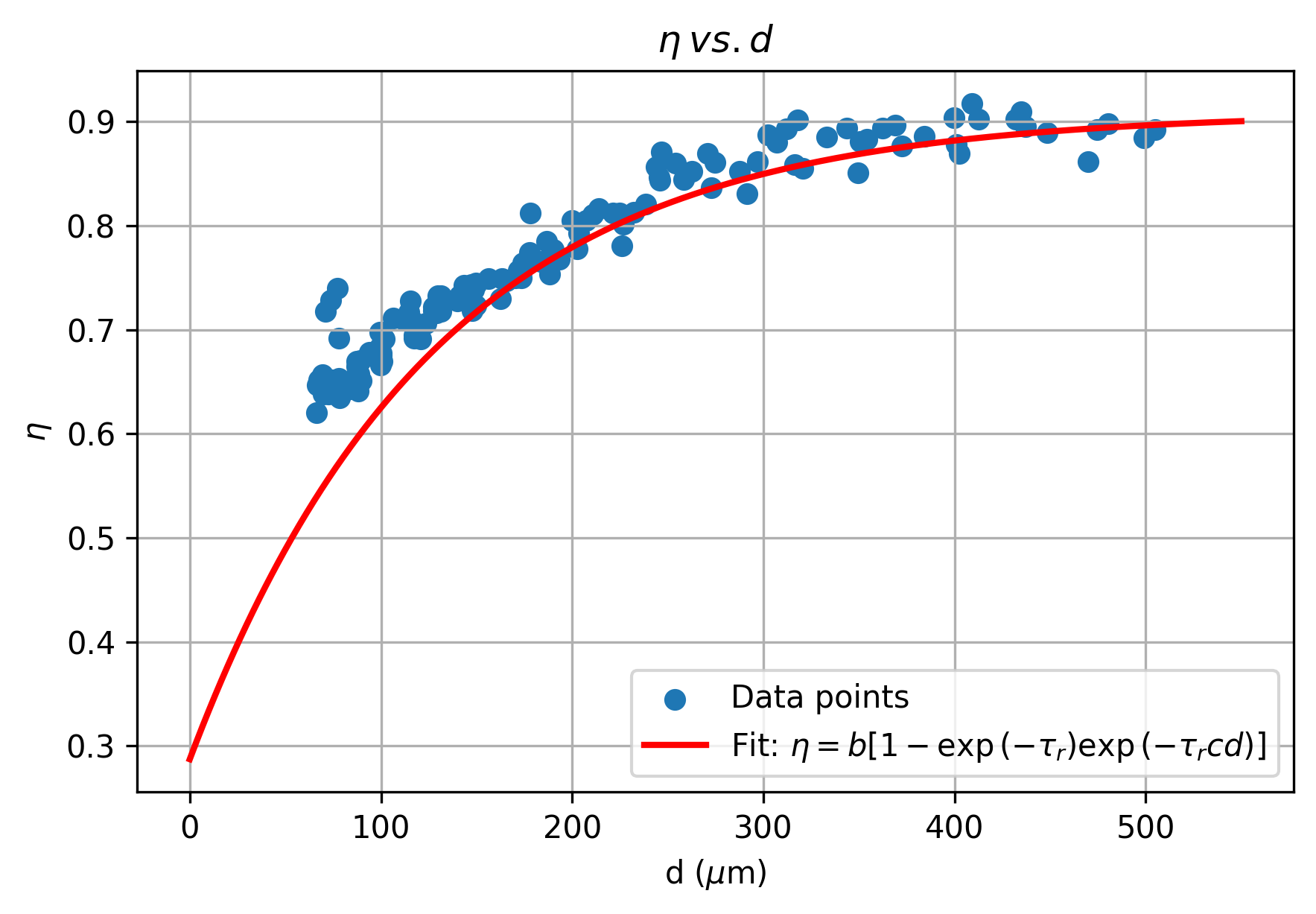}
        \caption{Alloy~718, $\phi=77~\mu\mathrm{m}$}
        \label{subfig:sim_etavsd_IN-77}
    \end{subfigure}
    \hfill
    \begin{subfigure}[b]{0.32\textwidth}
        \centering
        \includegraphics[width=\textwidth]{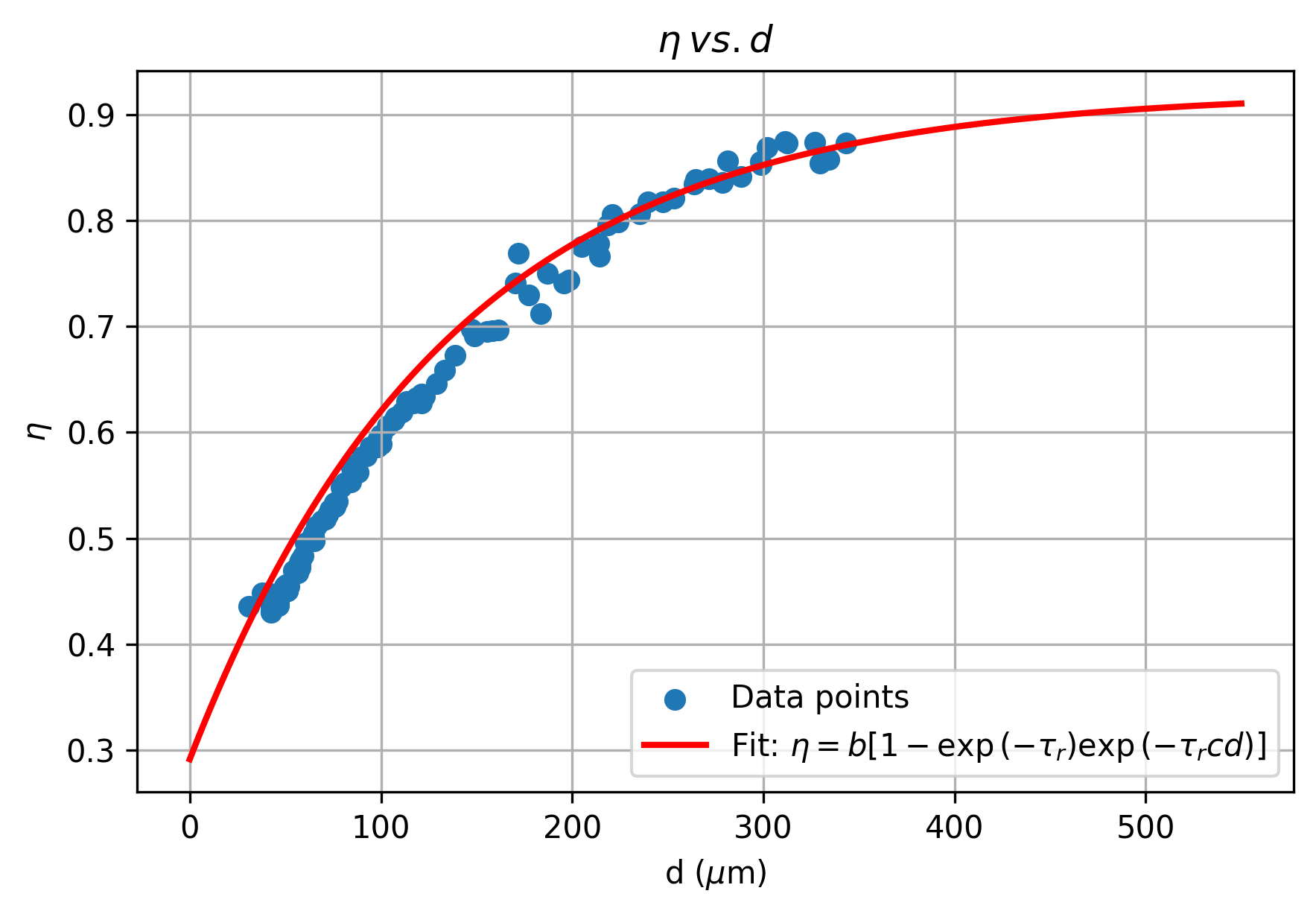}
        \caption{Alloy~718, $\phi=100~\mu\mathrm{m}$}
        \label{subfig:sim_etavsd_IN-100}
    \end{subfigure}
    \hfill
    \begin{subfigure}[b]{0.32\textwidth}
        \centering
        \includegraphics[width=\textwidth]{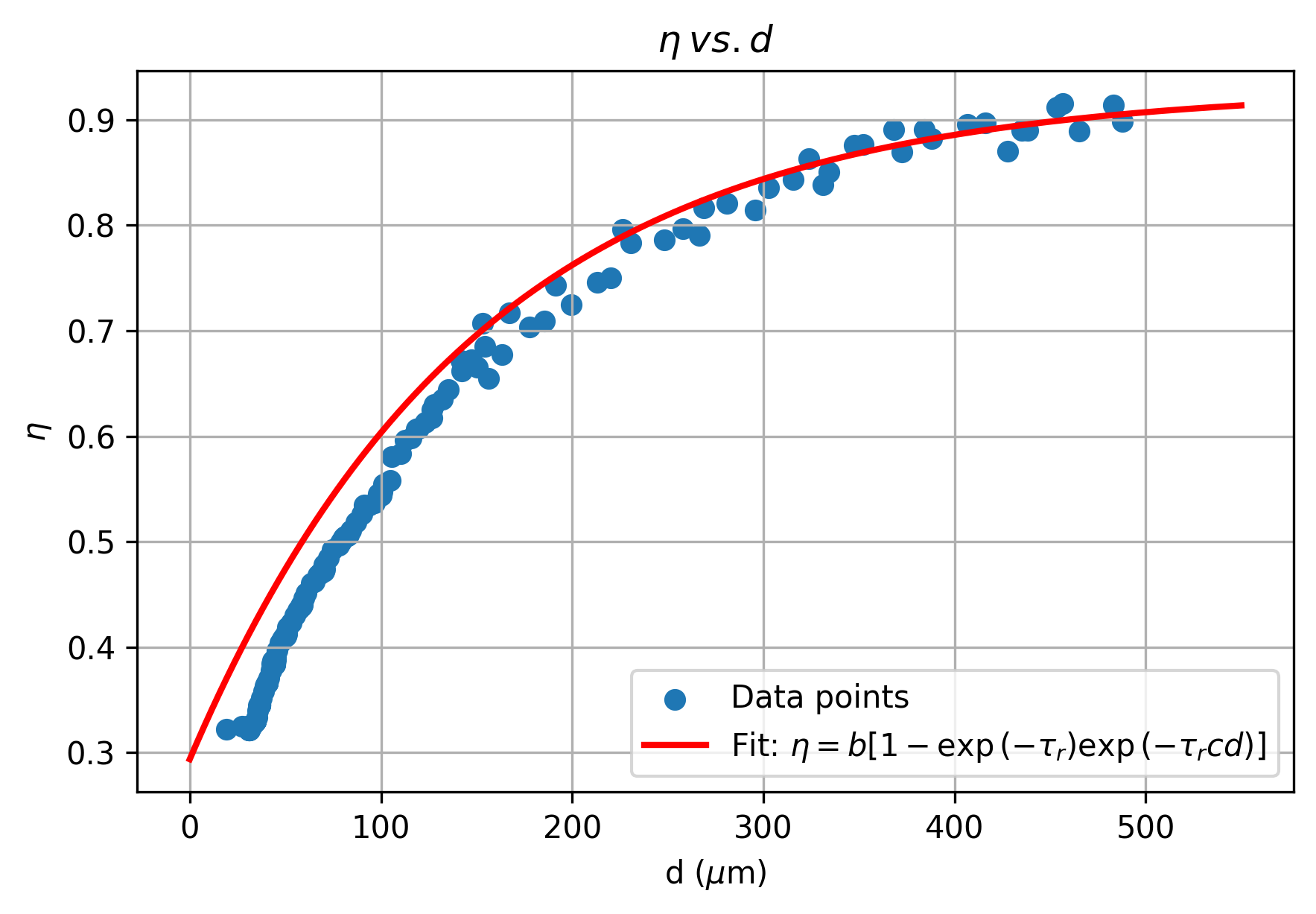}
        \caption{Alloy~718, $\phi=123~\mu\mathrm{m}$}
        \label{subfig:sim_etavsd_IN-123}
    \end{subfigure}

    \vspace{0.3cm}

    \begin{subfigure}[b]{0.32\textwidth}
        \centering
        \includegraphics[width=\textwidth]{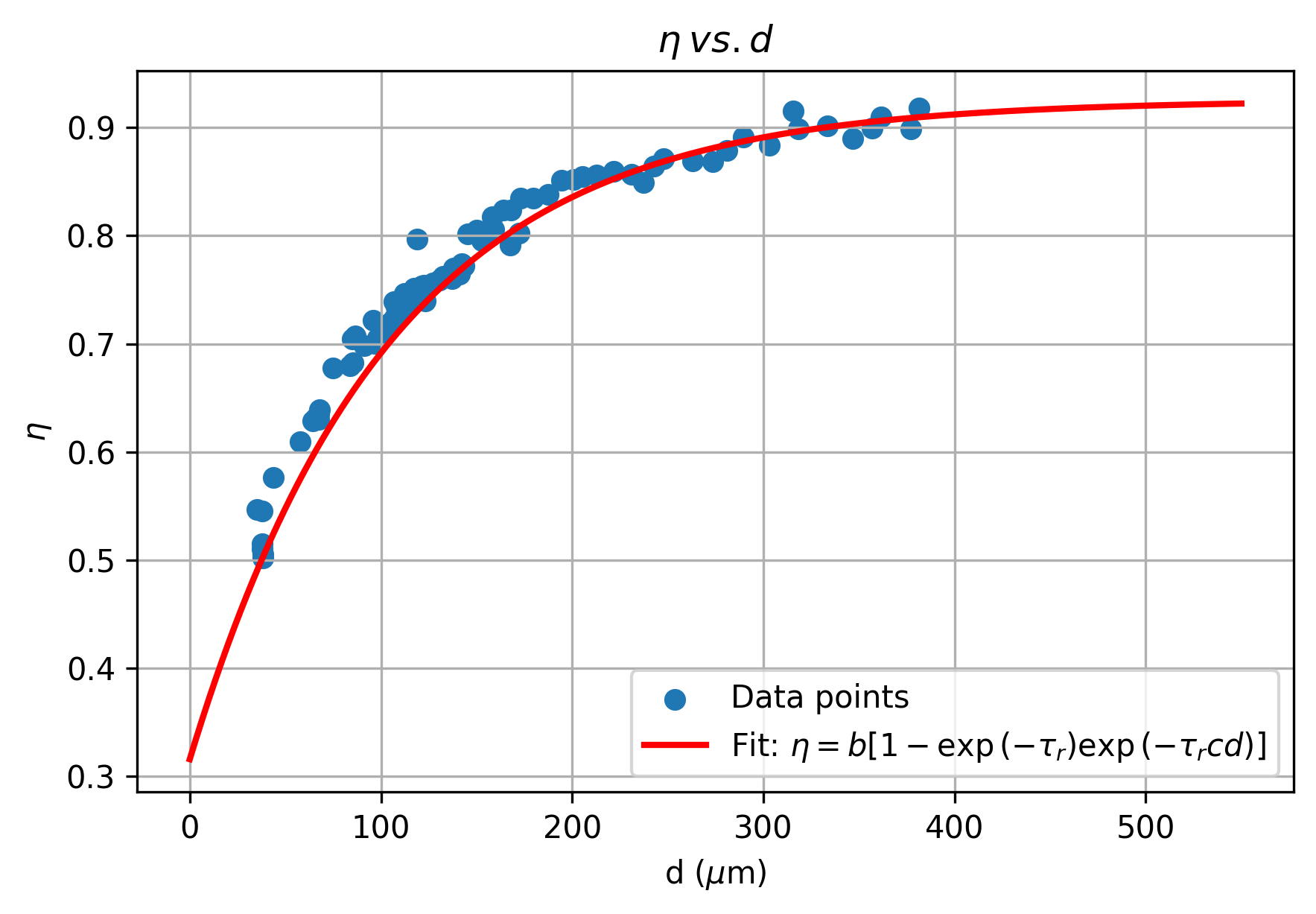}
        \caption{SS316, $\phi=77~\mu\mathrm{m}$}
        \label{subfig:sim_etavsd_SS316-77}
    \end{subfigure}
    \hfill
    \begin{subfigure}[b]{0.32\textwidth}
        \centering
        \includegraphics[width=\textwidth]{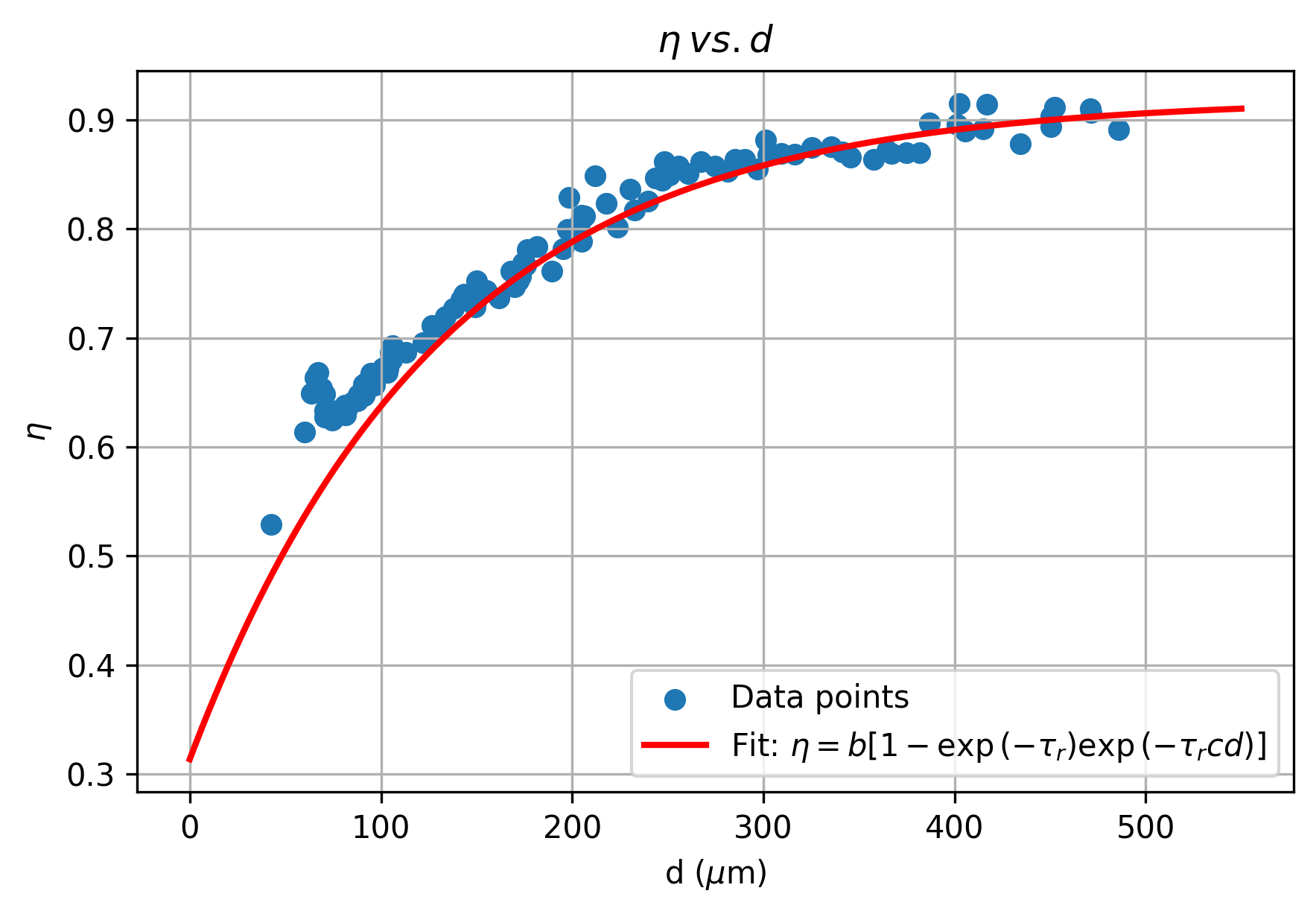}
        \caption{SS316, $\phi=100~\mu\mathrm{m}$}
        \label{subfig:sim_etavsd_SS316-100}
    \end{subfigure}
    \hfill
    \begin{subfigure}[b]{0.32\textwidth}
        \centering
        \includegraphics[width=\textwidth]{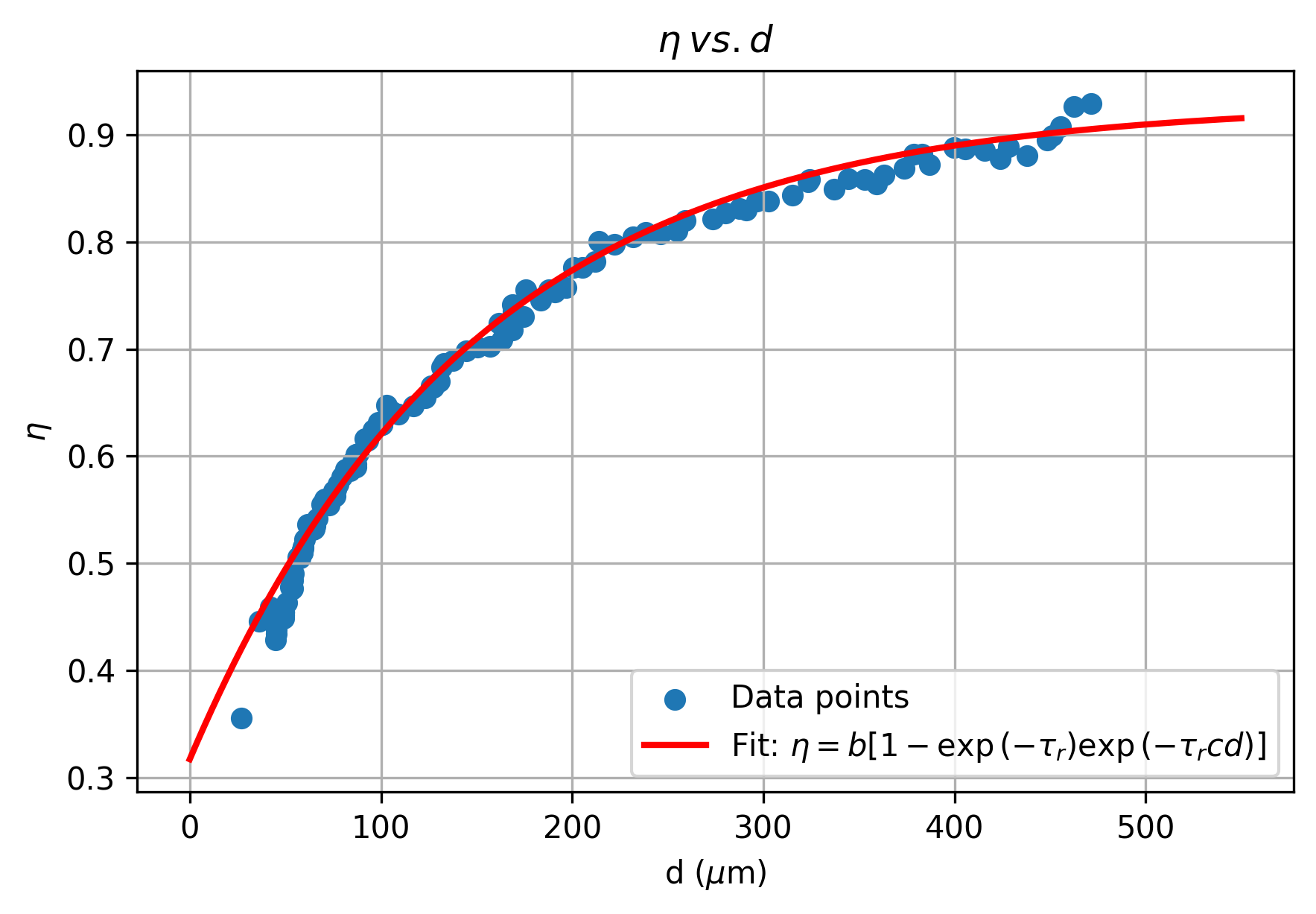}
        \caption{SS316, $\phi=123~\mu\mathrm{m}$}
        \label{subfig:sim_etavsd_SS316-123}
    \end{subfigure}
    
    \caption{Simulated total absorptance, $\eta$, as a function of cavity depth, $d$, for all nine material--beam groups. Blue points denote individual cavity states, and red curves show Eq.~\eqref{eqn:eta_3} with EPROD prescribed, $c$ fixed at the independently fitted value in Tab.~\ref{tab:sim_Nvsd}, and only $b$ fitted. Rows show Ti-64, Alloy~718, and SS316; columns show $\phi=77$, $100$, and $123~\mu\mathrm{m}$.}
    \label{fig:sim_etavsd}
\end{figure*}

With $c$ fixed at the independently identified values in Tab.~\ref{tab:sim_Nvsd}, Eq.~\eqref{eqn:eta_3} reproduces the dominant monotonic and saturating absorptance trend across the nine simulation groups (Fig.~\ref{fig:sim_etavsd}). Fitting only $b$ gives $R^2=0.6559$--$0.9863$ and MAPE$=2.01\%$--$12.61\%$, with a median MAPE of $2.61\%$. Six of the nine conditions achieve $R^2\geq0.9030$, while the MAPE values provide a complementary measure for conditions spanning comparatively narrow absorptance ranges.

\begin{table}[!htbp]
\centering
\caption{Fits of Eq.~\eqref{eqn:eta_3} to simulated total absorptance for each material and laser spot diameter $\phi$. For every condition, EPROD is prescribed, and $c$ is fixed at the corresponding value in Tab.~\ref{tab:sim_Nvsd}; only $b$ is fitted.}
\label{tab:sim_etavsd}
\begingroup
\small
\renewcommand{\arraystretch}{1.20}
\setlength{\tabcolsep}{3.5pt}

\sisetup{
    detect-weight=true,
    detect-family=true,
    table-number-alignment=center,
    table-text-alignment=center
}

\begin{tabular*}{\linewidth}{
    @{\hspace{0.6em}\extracolsep{\fill}}
    l
    *{3}{
        S[table-format=1.4]
        S[table-format=1.4]
        S[table-format=2.2]
    }
    @{\hspace{0.6em}}
}
\toprule

\multicolumn{1}{c}{%
    \multirow[c]{3}{*}{%
        \makecell[c]{\bfseries\boldmath
        $d$~\textit{vs.}~$\eta$\\[-0.2ex]simulation}}}
& \multicolumn{3}{c}{$\phi=77~\mu\mathrm{m}$}
& \multicolumn{3}{c}{$\phi=100~\mu\mathrm{m}$}
& \multicolumn{3}{c}{$\phi=123~\mu\mathrm{m}$} \\
\cmidrule(lr){2-4}
\cmidrule(lr){5-7}
\cmidrule(lr){8-10}

& \multicolumn{1}{c}{\multirow[c]{2}{*}{$b$}}
& \multicolumn{1}{c}{\multirow[c]{2}{*}{$R^2$}}
& \multicolumn{1}{c}{MAPE}
& \multicolumn{1}{c}{\multirow[c]{2}{*}{$b$}}
& \multicolumn{1}{c}{\multirow[c]{2}{*}{$R^2$}}
& \multicolumn{1}{c}{MAPE}
& \multicolumn{1}{c}{\multirow[c]{2}{*}{$b$}}
& \multicolumn{1}{c}{\multirow[c]{2}{*}{$R^2$}}
& \multicolumn{1}{c}{MAPE} \\[-0.2ex]

&
&
& \multicolumn{1}{c}{(\%)}
&
&
& \multicolumn{1}{c}{(\%)}
&
&
& \multicolumn{1}{c}{(\%)} \\

\midrule

\makecell[l]{Ti-64\\[-0.2ex]($\varepsilon=0.385$)}
& 0.9337
& 0.9059
& 2.31
& 0.9321
& 0.7043
& 2.61
& 0.9278
& 0.9825
& 2.01 \\
\addlinespace[2pt]

\makecell[l]{Alloy~718\\[-0.2ex]($\varepsilon=0.316$)}
& 0.9089
& 0.6559
& 5.42
& 0.9216
& 0.9668
& 4.43
& 0.9299
& 0.9030
& 12.61 \\
\addlinespace[2pt]

\makecell[l]{SS316\\[-0.2ex]($\varepsilon=0.341$)}
& 0.9255
& 0.9487
& 2.61
& 0.9195
& 0.8745
& 3.89
& 0.9298
& 0.9863
& 2.37 \\

\bottomrule
\end{tabular*}
\endgroup
\end{table}

The fitted closure factor remains within $b=0.9089$--$0.9337$, with a cross-condition mean and sample standard deviation of $0.9254\pm0.0078$. Its narrow distribution demonstrates that, after the EPROD-independent geometric scale $c$ is identified from $\overline{N}_w$, the remaining distributional and optical contributions can be represented by a compact and stable multiplicative closure across the examined material--beam matrix. Thus, the explicit material dependence is carried primarily by EPROD, while $b$ accounts for the residual difference between the leading mean-count representation and the complete beam-level absorptance.

\subsection{Evaluation using synchronized synchrotron-imaging experimental data}
\label{subsec:experiment_results}
\begin{figure}[!htbp]
    \centering
    \includegraphics[width=0.8\textwidth]{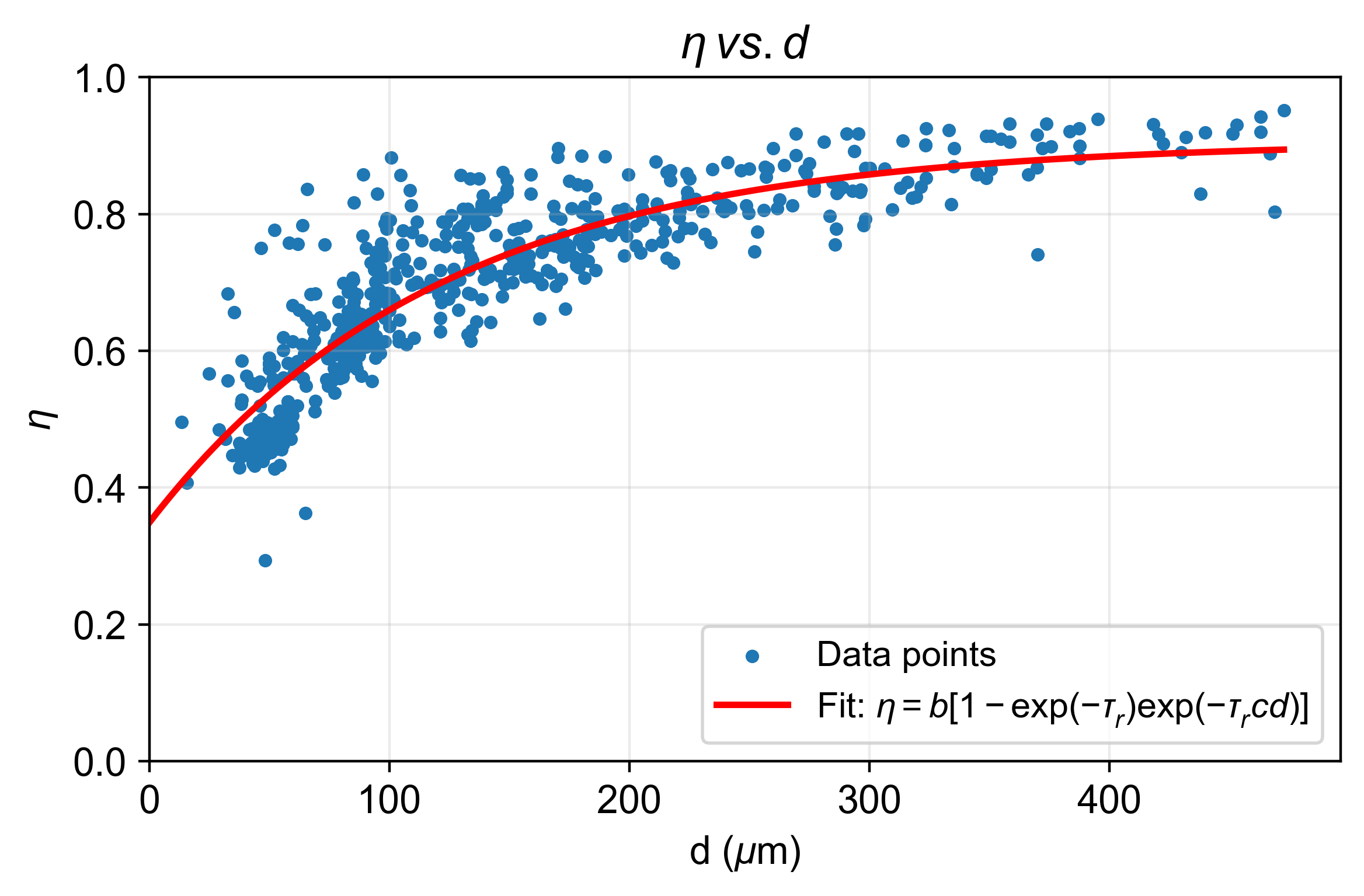}
    \caption{Synchronized total absorptance and X-ray-derived cavity depth for Ti-64 irradiated with a $122.5\pm3~\mu\mathrm{m}$ Gaussian laser spot~\cite{SIMONDS2020775}. Blue points denote experimental $\left(d,\eta\right)$ pairs, and the red curve shows the fit to Eq.~\eqref{eqn:eta_3} with $\varepsilon=0.385$ ($\mathrm{EPROD}=0.4861$), yielding $c=0.0168~\mu\mathrm{m}^{-1}$ and $b=0.9059$.}
    \label{fig:exp_etavsd}
\end{figure}

We next evaluated the composite absorptance--depth relation using the synchronized Ti-64 measurements reported by Simonds~\textit{et~al.}~\cite{SIMONDS2020775,simonds2021causal}. The experiment combines integrating-sphere radiometry with high-speed synchrotron X-ray imaging, providing directly paired measurements of $\eta$ and $d$ for a Gaussian laser spot diameter of $122.5\pm3~\mu\mathrm{m}$. Fitting Eq.~\eqref{eqn:eta_3} to the measured $(d,\eta)$ pairs with $\varepsilon=0.385$ and $\tau_r=0.4861$ prescribed yields the relation shown in Fig.~\ref{fig:exp_etavsd}, which closely captures both the increase in absorptance with keyhole depth and its gradual saturation over measured depths extending to approximately $480~\mu\mathrm{m}$.

The experimental fit yields $c=0.0168~\mu\mathrm{m}^{-1}$ and $b=0.9059$. These values closely agree with the corresponding Ti-64 simulation results of $c=0.0164~\mu\mathrm{m}^{-1}$ and $b=0.9278$ at the nearly identical spot diameter of $123~\mu\mathrm{m}$, differing by only approximately $2.4\%$ in each case. The corresponding reciprocal depth scales are approximately $60$ and $61~\mu\mathrm{m}$, respectively. It is worth noting that the simulated $c$ was obtained directly from ray-resolved $\overline{N}_w$--$d$ statistics and subsequently fixed in the absorptance fit, whereas the experimental $c$ and $b$ were jointly inferred from the measured absorptance--depth relation without constraints from the simulation. Because ray-resolved interaction statistics are unavailable experimentally, the fitted experimental $c$ is an effective coefficient of the composite absorptance--depth relation rather than a direct measurement of $\overline{N}_w$. Nevertheless, its close agreement with the independently obtained mean-count coefficient supports the use of $\overline{N}_w$ to identify the characteristic geometric depth scale in the reduced-order model.

\subsection{Discussion}
\label{subsec:discussion}
The reduced-order framework converts the coupled geometry--absorptance response into three physically interpretable components. EPROD describes the prescribed optical attenuation per laser--wall interaction, $c$ characterizes the depthwise accumulation of power-weighted mean interactions, and $b$ provides the beam-level closure and sets the asymptotic absorptance. The fitted $c$ values correspond to one additional mean interaction per approximately $43$--$61~\mu\mathrm{m}$ of cavity depth, while $(\tau_r c)^{-1}\approx100$--$150~\mu\mathrm{m}$ defines the associated optical--geometric attenuation length. The substantially smaller cross-alloy variation in $c$ than its systematic variation with beam diameter supports its interpretation as an approximately alloy-stable leading scale within the examined matrix. At the same time, the non-constant values of $c\phi$ indicate that spot diameter alone does not fully organize the ray trajectories; opening geometry, wall inclination, curvature, and three-dimensional asymmetry remain relevant. Thus, $c$ is best interpreted as a transferable, beam- and morphology-conditioned interaction density rather than either a universal constant or a purely empirical fit parameter.

The finite-beam statistics also distinguish the statewise exactness of $N_{\mathrm{eff}}$ from the accuracy of a depth-based linear approximation to it. For a prescribed constant EPROD, $N_{\mathrm{eff}}$ exactly aggregates the complete power-weighted ray-count distribution at each cavity state. This exactness does not carry over when $N_{\mathrm{eff}}(d)$ is compressed into a unit-intercept form $1+c_{\mathrm{eff}}d$, because the depth-dependent difference $\Delta(d)=\overline{N}_w-N_{\mathrm{eff}}$ is then represented by a single slope. Moreover, a slope identified in interaction-count space need not minimize the error after the nonlinear mapping to absorptance, and the full ray-traced absorptance additionally retains incidence-angle-dependent Fresnel effects absent from the constant-EPROD reference. A coefficient fitted to $N_{\mathrm{eff}}$ therefore need not outperform the $\overline{N}_w$-derived $c$ in the present two-parameter closure. Rather, $\overline{N}_w$ supplies the EPROD-independent leading geometric scale, whereas $N_{\mathrm{eff}}$ retains the material-conditioned influence of the complete ray-history distribution. Because $\Delta(d)$ evolves with cavity depth, a straight line fitted only to finite-depth $N_{\mathrm{eff}}$ data also need not extrapolate through $(0,1)$; such an intercept mismatch is not, by itself, evidence that the constant-EPROD representation has failed.

Within the present simulations, the second-order cumulant correction is generally about $15\%$ or less for $d\lesssim100~\mu\mathrm{m}$, identifying a shallow-cavity range in which $N_{\mathrm{eff}}\approx\overline{N}_w$ is a useful leading approximation. The correction becomes appreciable over the condition-dependent intermediate range of approximately $150$--$350~\mu\mathrm{m}$, where its clustered maxima indicate the coexistence of short escape paths and more strongly confined ray families. At greater depths, the normalized correction generally weakens, although the two interaction-count measures need not become identical. When the $\overline{N}_w$-derived depth scale is propagated through the absorptance closure, the fitted $b$ absorbs the average residual distributional and optical contributions, while exponential attenuation compresses count-space deviations as the response approaches saturation. This mechanism explains why the global absorptance relation can remain accurate despite statewise departures from the leading mean-count trend. The tightly clustered $b$ values and the close experimental--simulation agreement under matched beam geometry together provide an independent consistency check of this composite parameterization.

The constant material-specific EPROD used to calculate $N_{\mathrm{eff}}$ should be viewed as a controlled reference that enables geometric interaction accumulation to be separated from optical attenuation. The full ray-tracing calculation retains the local incidence-angle dependence of Fresnel absorption, while the actual optical response may additionally evolve with temperature and surface state. A natural extension is therefore to replace $\tau_rN$ with the path-integrated optical depth accumulated along each ray. This extension would retain the present statistical hierarchy while accommodating more strongly varying material and interfacial conditions.

Beyond its physical interpretation, Eq.~\eqref{eqn:eta_3} provides an inexpensive energy-coupling closure for reduced-order thermal-fluid models, process maps, and time-resolved monitoring. Given absorptance, the relation can be inverted to estimate cavity depth as
\begin{equation}
d(\eta)
=
\frac{-\ln\!\left(1-\eta/b\right)/\tau_r-1}{c}
\end{equation}
The corresponding depth sensitivity is
\begin{equation}
\frac{\partial\eta}{\partial d}
=
b\tau_r c\exp\!\left[-\tau_r(1+cd)\right]
\end{equation}
For positive $b$, $\tau_r$, and $c$, this sensitivity decreases monotonically with increasing depth. Absorptance-based depth estimation is therefore best conditioned toward the shallow end of the keyhole regime and becomes progressively less sensitive with increasing depth, particularly as $\eta$ approaches $b$. Persistent departures from the calibrated relation may additionally indicate changes in cavity morphology or reorganizations among ray-path families. The framework consequently provides a practical hierarchy from rapid depth-only prediction to distribution-informed analysis and fully ray-resolved optical modeling.

\section{Conclusions and future work}
\label{sec:conclusion}
This work establishes a reduced-order connection among finite-beam ray statistics, vaporization-cavity depth, and total laser absorptance. Under a constant effective single-interaction absorptance, repeated laser--wall interactions yield a Beer--Lambert-type attenuation relation characterized by EPROD, while the finite-beam formulation distinguishes the exact absorptance-equivalent count from its leading power-weighted mean statistic. Across nine simulated material--beam groups, the unit-intercept relation $\overline{N}_w=1+cd$ captures the dominant depth dependence of the EPROD-independent mean interaction count. The fitted $c$ varies only weakly across alloys at matched spot diameter but decreases systematically with increasing beam diameter, identifying it as an approximately alloy-stable, beam- and morphology-conditioned interaction density within the examined matrix. The simultaneously evaluated $N_{\mathrm{eff}}$ values reveal the expected material- and distribution-dependent reduction from $\overline{N}_w$ and explain why the effective optical statistic can vary more smoothly than the underlying arithmetic mean.

With $c$ fixed independently from the $\overline{N}_w$ statistics and only $b$ fitted, the absorptance--depth closure achieves a median MAPE of $2.61\%$, while $b$ remains tightly concentrated within $0.9089$--$0.9337$. A synchronized Ti-64 experiment under nearly matched beam geometry independently yields $c=0.0168~\mu\mathrm{m}^{-1}$ and $b=0.9059$, each within approximately $2.4\%$ of the corresponding simulation value. The close agreement between the independently obtained experimental and simulation parameters supports the reproducibility of the characteristic depth and absorptance-closure scales under nearly matched beam geometry. Separately, the recurring magnitude of the variance correction across the nine simulation conditions reveals a common statistical structure in the reorganization of finite-beam ray trajectories.

The resulting framework provides a computationally efficient bridge from cavity geometry to global absorbed power and establishes a hierarchy ranging from rapid depth-only prediction to distribution-informed and fully ray-resolved descriptions. Future studies can extend this hierarchy through path-integrated optical depths, held-out transfer tests, additional beam profiles and material systems, and compact morphology descriptors that complement cavity depth. These developments can further support reduced-order thermal modeling, absorptance-based keyhole tracking, and physics-informed monitoring of laser--metal interactions.

\section*{Acknowledgement}
\label{sec:acknow}
This work was supported in part by the National Science Foundation under Grant No.~{CMMI-2412545}. We thank Dr. James Fraser for helpful discussions. Haolin Liu thanks Silin Liu for assistance with manuscript proofreading and figure preparation. 

\textbf{CRediT authorship contribution statement.} Haolin Liu: Conceptualization, Methodology, Formal analysis, Software, Validation, Data curation, Visualization, Writing--original draft, Writing--review \& editing, Project administration. Haoran Shi: Methodology, Software, Investigation, Data curation, Validation, Writing--review \& editing. Xuzhe Zeng: Software, Investigation, Data analysis, Validation, Writing--review \& editing. Brian Simonds: Investigation, Resources, Data curation, Writing--review \& editing. Wenda Tan: Methodology, Resources, Supervision, Writing--review \& editing. Anthony D. Rollett: Resources, Supervision, Funding acquisition, Writing--review \& editing. Levent Burak Kara: Supervision, Project administration, Funding acquisition, Writing--review \& editing.

\textbf{Declaration of generative AI and AI-assisted technologies in the manuscript preparation process.} During the preparation of this work, the lead author used ChatGPT (OpenAI), Gemini (Google), and Claude (Anthropic) as supporting tools for theoretical discussions, code development and testing, and language editing and proofreading. All AI-generated outputs were critically reviewed, independently verified where applicable, and revised before being incorporated into the work. All authors reviewed and approved the final manuscript and take full responsibility for its content.

\bibliographystyle{elsarticle-num-names}
\bibliography{references}

\begin{thebibliography}{28}
\expandafter\ifx\csname natexlab\endcsname\relax\def\natexlab#1{#1}\fi
\providecommand{\url}[1]{\texttt{#1}}
\providecommand{\href}[2]{#2}
\providecommand{\path}[1]{#1}
\providecommand{\DOIprefix}{doi:}
\providecommand{\ArXivprefix}{arXiv:}
\providecommand{\URLprefix}{URL: }
\providecommand{\Pubmedprefix}{pmid:}
\providecommand{\doi}[1]{\href{http://dx.doi.org/#1}{\path{#1}}}
\providecommand{\Pubmed}[1]{\href{pmid:#1}{\path{#1}}}
\providecommand{\bibinfo}[2]{#2}
\ifx\xfnm\relax \def\xfnm[#1]{\unskip,\space#1}\fi
\bibitem[{Fabbro et~al.(2005)Fabbro, Slimani, Coste, and Briand}]{Fabbro2005Study}
\bibinfo{author}{R.~Fabbro}, \bibinfo{author}{S.~Slimani}, \bibinfo{author}{F.~Coste}, \bibinfo{author}{F.~Briand},
\newblock \bibinfo{title}{Study of keyhole behaviour for full penetration nd-yag cw laser welding},
\newblock \bibinfo{journal}{Journal of Physics D: Applied Physics} \bibinfo{volume}{38} (\bibinfo{year}{2005}) \bibinfo{pages}{1881--1887}.
\bibitem[{King et~al.(2015)King, Anderson, Ferencz, Hodge, Kamath, Khairallah, and Rubenchik}]{king2015laser}
\bibinfo{author}{W.~E. King}, \bibinfo{author}{A.~T. Anderson}, \bibinfo{author}{R.~M. Ferencz}, \bibinfo{author}{N.~E. Hodge}, \bibinfo{author}{C.~Kamath}, \bibinfo{author}{S.~A. Khairallah}, \bibinfo{author}{A.~M. Rubenchik},
\newblock \bibinfo{title}{Laser powder bed fusion additive manufacturing of metals; physics, computational, and materials challenges},
\newblock \bibinfo{journal}{Applied Physics Reviews} \bibinfo{volume}{2} (\bibinfo{year}{2015}) \bibinfo{pages}{041304}.
\bibitem[{Khairallah et~al.(2016)Khairallah, Anderson, Rubenchik, and King}]{khairallah2016laser}
\bibinfo{author}{S.~A. Khairallah}, \bibinfo{author}{A.~T. Anderson}, \bibinfo{author}{A.~Rubenchik}, \bibinfo{author}{W.~E. King},
\newblock \bibinfo{title}{Laser powder-bed fusion additive manufacturing: Physics of complex melt flow and formation mechanisms of pores, spatter, and denudation zones},
\newblock \bibinfo{journal}{Acta Materialia} \bibinfo{volume}{108} (\bibinfo{year}{2016}) \bibinfo{pages}{36--45}.
\bibitem[{Cunningham et~al.(2019)Cunningham, Zhao, Parab, Kantzos, Pauza, Fezzaa, Sun, and Rollett}]{cunningham2019keyhole}
\bibinfo{author}{R.~Cunningham}, \bibinfo{author}{C.~Zhao}, \bibinfo{author}{N.~Parab}, \bibinfo{author}{C.~Kantzos}, \bibinfo{author}{J.~Pauza}, \bibinfo{author}{K.~Fezzaa}, \bibinfo{author}{T.~Sun}, \bibinfo{author}{A.~D. Rollett},
\newblock \bibinfo{title}{Keyhole threshold and morphology in laser melting revealed by ultrahigh-speed x-ray imaging},
\newblock \bibinfo{journal}{Science} \bibinfo{volume}{363} (\bibinfo{year}{2019}) \bibinfo{pages}{849--852}.
\bibitem[{Zhang et~al.(2013)Zhang, Chen, Zhou, and Li}]{zhang2013direct}
\bibinfo{author}{M.~Zhang}, \bibinfo{author}{G.~Chen}, \bibinfo{author}{Y.~Zhou}, \bibinfo{author}{S.~Li},
\newblock \bibinfo{title}{Direct observation of keyhole characteristics in deep penetration laser welding with a 10 kw fiber laser},
\newblock \bibinfo{journal}{Optics Express} \bibinfo{volume}{21} (\bibinfo{year}{2013}) \bibinfo{pages}{19997--20004}.
\bibitem[{Kaplan(2012)}]{kaplan2012fresnel}
\bibinfo{author}{A.~F.~H. Kaplan},
\newblock \bibinfo{title}{Fresnel absorption of 1 $\mu$m- and 10 $\mu$m-laser beams at the keyhole wall during laser beam welding: Comparison between smooth and wavy surfaces},
\newblock \bibinfo{journal}{Applied Surface Science} \bibinfo{volume}{258} (\bibinfo{year}{2012}) \bibinfo{pages}{3354--3363}.
\bibitem[{Martin et~al.(2019)Martin, Calta, Khairallah, Wang, Depond, Fong, Thampy, Guss, Kiss, Stone, Tassone, Weker, Toney, van Buuren, and Matthews}]{martin2019dynamics}
\bibinfo{author}{A.~A. Martin}, \bibinfo{author}{N.~P. Calta}, \bibinfo{author}{S.~A. Khairallah}, \bibinfo{author}{J.~Wang}, \bibinfo{author}{P.~J. Depond}, \bibinfo{author}{A.~Y. Fong}, \bibinfo{author}{V.~Thampy}, \bibinfo{author}{G.~M. Guss}, \bibinfo{author}{A.~M. Kiss}, \bibinfo{author}{K.~H. Stone}, \bibinfo{author}{C.~J. Tassone}, \bibinfo{author}{J.~N. Weker}, \bibinfo{author}{M.~F. Toney}, \bibinfo{author}{T.~van Buuren}, \bibinfo{author}{M.~J. Matthews},
\newblock \bibinfo{title}{Dynamics of pore formation during laser powder bed fusion additive manufacturing},
\newblock \bibinfo{journal}{Nature Communications} \bibinfo{volume}{10} (\bibinfo{year}{2019}) \bibinfo{pages}{1987}.
\bibitem[{Deisenroth et~al.(2026)Deisenroth, Shi, Wittermer, Valiyakath Vadakkan~Habeeb, Chou, Tan, Weaver, and Wudy}]{deisenroth2026illuminating}
\bibinfo{author}{D.~Deisenroth}, \bibinfo{author}{H.~Shi}, \bibinfo{author}{M.~Wittermer}, \bibinfo{author}{N.~Valiyakath Vadakkan~Habeeb}, \bibinfo{author}{K.~Chou}, \bibinfo{author}{W.~Tan}, \bibinfo{author}{J.~Weaver}, \bibinfo{author}{K.~Wudy},
\newblock \bibinfo{title}{Illuminating the physics of melting during laser-based manufacturing of in718 by measuring laser light reflections},
\newblock \bibinfo{journal}{Additive Manufacturing}  (\bibinfo{year}{2026}) \bibinfo{pages}{105330}. \DOIprefix\doi{10.1016/j.addma.2026.105330}.
\bibitem[{Zhao et~al.(2022)Zhao, Shi, Chen, Du, Sun, Simonds, Fezzaa, and Rollett}]{zhao2022laser}
\bibinfo{author}{C.~Zhao}, \bibinfo{author}{B.~Shi}, \bibinfo{author}{S.~Chen}, \bibinfo{author}{D.~Du}, \bibinfo{author}{T.~Sun}, \bibinfo{author}{B.~J. Simonds}, \bibinfo{author}{K.~Fezzaa}, \bibinfo{author}{A.~D. Rollett},
\newblock \bibinfo{title}{Laser melting modes in metal powder bed fusion additive manufacturing},
\newblock \bibinfo{journal}{Reviews of Modern Physics} \bibinfo{volume}{94} (\bibinfo{year}{2022}) \bibinfo{pages}{045002}.
\bibitem[{Trapp et~al.(2017)Trapp, Rubenchik, Guss, and Matthews}]{trapp2017situ}
\bibinfo{author}{J.~Trapp}, \bibinfo{author}{A.~M. Rubenchik}, \bibinfo{author}{G.~Guss}, \bibinfo{author}{M.~J. Matthews},
\newblock \bibinfo{title}{In situ absorptivity measurements of metallic powders during laser powder-bed fusion additive manufacturing},
\newblock \bibinfo{journal}{Applied Materials Today} \bibinfo{volume}{9} (\bibinfo{year}{2017}) \bibinfo{pages}{341--349}.
\bibitem[{Simonds et~al.(2018)Simonds, Sowards, Hadler, Pfeif, Wilthan, Tanner, Harris, Williams, and Lehman}]{simonds2018time}
\bibinfo{author}{B.~J. Simonds}, \bibinfo{author}{J.~Sowards}, \bibinfo{author}{J.~Hadler}, \bibinfo{author}{E.~Pfeif}, \bibinfo{author}{B.~Wilthan}, \bibinfo{author}{J.~Tanner}, \bibinfo{author}{C.~Harris}, \bibinfo{author}{P.~Williams}, \bibinfo{author}{J.~Lehman},
\newblock \bibinfo{title}{Time-resolved absorptance and melt pool dynamics during intense laser irradiation of a metal},
\newblock \bibinfo{journal}{Physical review applied} \bibinfo{volume}{10} (\bibinfo{year}{2018}) \bibinfo{pages}{044061}.
\bibitem[{Ye et~al.(2019)Ye, Khairallah, Rubenchik, Crumb, Guss, Belak, and Matthews}]{ye2019energy}
\bibinfo{author}{J.~Ye}, \bibinfo{author}{S.~A. Khairallah}, \bibinfo{author}{A.~M. Rubenchik}, \bibinfo{author}{M.~F. Crumb}, \bibinfo{author}{G.~Guss}, \bibinfo{author}{J.~Belak}, \bibinfo{author}{M.~J. Matthews},
\newblock \bibinfo{title}{Energy coupling mechanisms and scaling behavior associated with laser powder bed fusion additive manufacturing},
\newblock \bibinfo{journal}{Advanced Engineering Materials} \bibinfo{volume}{21} (\bibinfo{year}{2019}) \bibinfo{pages}{1900185}.
\bibitem[{Allen et~al.(2020)Allen, Huang, Tanner, Tan, Fraser, and Simonds}]{allen2020energy}
\bibinfo{author}{T.~R. Allen}, \bibinfo{author}{W.~Huang}, \bibinfo{author}{J.~R. Tanner}, \bibinfo{author}{W.~Tan}, \bibinfo{author}{J.~M. Fraser}, \bibinfo{author}{B.~J. Simonds},
\newblock \bibinfo{title}{Energy-coupling mechanisms revealed through simultaneous keyhole depth and absorptance measurements during laser-metal processing},
\newblock \bibinfo{journal}{Physical Review Applied} \bibinfo{volume}{13} (\bibinfo{year}{2020}) \bibinfo{pages}{064070}.
\bibitem[{Cho and Na(2006)}]{cho2006implementation}
\bibinfo{author}{J.-H. Cho}, \bibinfo{author}{S.-J. Na},
\newblock \bibinfo{title}{Implementation of real-time multiple reflection and fresnel absorption of laser beam in keyhole},
\newblock \bibinfo{journal}{Journal of Physics D: Applied Physics} \bibinfo{volume}{39} (\bibinfo{year}{2006}) \bibinfo{pages}{5372--5378}.
\bibitem[{Kouraytem et~al.(2019)Kouraytem, Li, Cunningham, Zhao, Parab, Sun, Rollett, Spear, and Tan}]{kouraytem2019effect}
\bibinfo{author}{N.~Kouraytem}, \bibinfo{author}{X.~Li}, \bibinfo{author}{R.~Cunningham}, \bibinfo{author}{C.~Zhao}, \bibinfo{author}{N.~Parab}, \bibinfo{author}{T.~Sun}, \bibinfo{author}{A.~D. Rollett}, \bibinfo{author}{A.~D. Spear}, \bibinfo{author}{W.~Tan},
\newblock \bibinfo{title}{Effect of laser-matter interaction on molten pool flow and keyhole dynamics},
\newblock \bibinfo{journal}{Physical Review Applied} \bibinfo{volume}{11} (\bibinfo{year}{2019}) \bibinfo{pages}{064054}.
\bibitem[{Jin et~al.(2012)Jin, Cheng, Zeng, Zou, and Zhang}]{jin2012multiple}
\bibinfo{author}{X.~Jin}, \bibinfo{author}{Y.~Cheng}, \bibinfo{author}{L.~Zeng}, \bibinfo{author}{Y.~Zou}, \bibinfo{author}{H.~Zhang},
\newblock \bibinfo{title}{Multiple reflections and fresnel absorption of gaussian laser beam in an actual 3d keyhole during deep-penetration laser welding},
\newblock \bibinfo{journal}{International Journal of Optics} \bibinfo{volume}{2012} (\bibinfo{year}{2012}) \bibinfo{pages}{361818}.
\bibitem[{Jiang et~al.(2024)Jiang, Smith, Yi, Sun, Simonds, and Rollett}]{jiang2024deep}
\bibinfo{author}{R.~Jiang}, \bibinfo{author}{J.~Smith}, \bibinfo{author}{Y.-T. Yi}, \bibinfo{author}{T.~Sun}, \bibinfo{author}{B.~J. Simonds}, \bibinfo{author}{A.~D. Rollett},
\newblock \bibinfo{title}{Deep learning approaches for instantaneous laser absorptance prediction in additive manufacturing},
\newblock \bibinfo{journal}{npj Computational Materials} \bibinfo{volume}{10} (\bibinfo{year}{2024}) \bibinfo{pages}{6}.
\bibitem[{Zhao and DebRoy(2003)}]{zhao2003macroporosity}
\bibinfo{author}{H.~Zhao}, \bibinfo{author}{T.~DebRoy},
\newblock \bibinfo{title}{Macroporosity free aluminum alloy weldments through numerical simulation of keyhole mode laser welding},
\newblock \bibinfo{journal}{Journal of Applied Physics} \bibinfo{volume}{93} (\bibinfo{year}{2003}) \bibinfo{pages}{10089--10096}.
\bibitem[{Rai et~al.(2007)Rai, Elmer, Palmer, and DebRoy}]{rai2007heat}
\bibinfo{author}{R.~Rai}, \bibinfo{author}{J.~Elmer}, \bibinfo{author}{T.~A. Palmer}, \bibinfo{author}{T.~DebRoy},
\newblock \bibinfo{title}{Heat transfer and fluid flow during keyhole mode laser welding of tantalum, ti--6al--4v, 304l stainless steel and vanadium},
\newblock \bibinfo{journal}{Journal of physics D: Applied physics} \bibinfo{volume}{40} (\bibinfo{year}{2007}) \bibinfo{pages}{5753--5766}.
\bibitem[{Tan et~al.(2013)Tan, Bailey, and Shin}]{tan2013investigation}
\bibinfo{author}{W.~Tan}, \bibinfo{author}{N.~Bailey}, \bibinfo{author}{Y.~Shin},
\newblock \bibinfo{title}{Investigation of keyhole plume and molten pool based on a three-dimensional dynamic model with sharp interface formulation},
\newblock \bibinfo{journal}{Journal of Physics D: Applied Physics} \bibinfo{volume}{46} (\bibinfo{year}{2013}) \bibinfo{pages}{055501}. \DOIprefix\doi{10.1088/0022-3727/46/5/055501}.
\bibitem[{Tan and Shin(2014)}]{tan2014analysis}
\bibinfo{author}{W.~Tan}, \bibinfo{author}{Y.~Shin},
\newblock \bibinfo{title}{Analysis of multi-phase interaction and its effects on keyhole dynamics with a multi-physics numerical model},
\newblock \bibinfo{journal}{Journal of Physics D: Applied Physics} \bibinfo{volume}{47} (\bibinfo{year}{2014}) \bibinfo{pages}{345501}. \DOIprefix\doi{10.1088/0022-3727/47/34/345501}.
\bibitem[{Li et~al.(2021)Li, Guo, Chen, and Tan}]{li2021quantitative}
\bibinfo{author}{X.~Li}, \bibinfo{author}{Q.~Guo}, \bibinfo{author}{L.~Chen}, \bibinfo{author}{W.~Tan},
\newblock \bibinfo{title}{Quantitative investigation of gas flow, powder-gas interaction, and powder behavior under different ambient pressure levels in laser powder bed fusion},
\newblock \bibinfo{journal}{International Journal of Machine Tools and Manufacture} \bibinfo{volume}{170} (\bibinfo{year}{2021}) \bibinfo{pages}{103797}. \DOIprefix\doi{10.1016/j.ijmachtools.2021.103797}.
\bibitem[{Polyanskiy(2024)}]{polyanskiy2024refractiveindex}
\bibinfo{author}{M.~N. Polyanskiy},
\newblock \bibinfo{title}{Refractiveindex.info database of optical constants},
\newblock \bibinfo{journal}{Scientific Data} \bibinfo{volume}{11} (\bibinfo{year}{2024}) \bibinfo{pages}{94}. \DOIprefix\doi{10.1038/s41597-023-02898-2}.
\bibitem[{Li et~al.(2020)Li, Zhao, Sun, and Tan}]{li2020revealing}
\bibinfo{author}{X.~Li}, \bibinfo{author}{C.~Zhao}, \bibinfo{author}{T.~Sun}, \bibinfo{author}{W.~Tan},
\newblock \bibinfo{title}{Revealing transient powder-gas interaction in laser powder bed fusion process through multi-physics modeling and high-speed synchrotron x-ray imaging},
\newblock \bibinfo{journal}{Additive Manufacturing} \bibinfo{volume}{35} (\bibinfo{year}{2020}) \bibinfo{pages}{101362}. \DOIprefix\doi{10.1016/j.addma.2020.101362}.
\bibitem[{Li et~al.(2025)Li, Bitharas, Moore, and Tan}]{li2025effects}
\bibinfo{author}{F.~Li}, \bibinfo{author}{I.~Bitharas}, \bibinfo{author}{A.~Moore}, \bibinfo{author}{W.~Tan},
\newblock \bibinfo{title}{Effects of gas composition and pressure level on powder spattering and denudation in laser powder bed fusion},
\newblock \bibinfo{journal}{Acta Materialia} \bibinfo{volume}{299} (\bibinfo{year}{2025}) \bibinfo{pages}{121443}. \DOIprefix\doi{10.1016/j.actamat.2025.121443}.
\bibitem[{Li et~al.(2026)Li, Shi, Kirschbaum, He, Okwudire, and Tan}]{li2026elucidating}
\bibinfo{author}{F.~Li}, \bibinfo{author}{H.~Shi}, \bibinfo{author}{N.~Kirschbaum}, \bibinfo{author}{C.~He}, \bibinfo{author}{C.~Okwudire}, \bibinfo{author}{W.~Tan},
\newblock \bibinfo{title}{Elucidating the mechanism for suppression of spatter in dual-laser powder bed fusion systems: A numerical and experimental study},
\newblock \bibinfo{journal}{Additive Manufacturing} \bibinfo{volume}{118} (\bibinfo{year}{2026}) \bibinfo{pages}{105102}. \DOIprefix\doi{10.1016/j.addma.2026.105102}.
\bibitem[{Simonds et~al.(2020)Simonds, Tanner, Artusio-Glimpse, Williams, Parab, Zhao, and Sun}]{SIMONDS2020775}
\bibinfo{author}{B.~J. Simonds}, \bibinfo{author}{J.~Tanner}, \bibinfo{author}{A.~Artusio-Glimpse}, \bibinfo{author}{P.~A. Williams}, \bibinfo{author}{N.~Parab}, \bibinfo{author}{C.~Zhao}, \bibinfo{author}{T.~Sun},
\newblock \bibinfo{title}{Simultaneous high-speed x-ray transmission imaging and absolute dynamic absorptance measurements during high-power laser-metal processing},
\newblock \bibinfo{journal}{Procedia CIRP} \bibinfo{volume}{94} (\bibinfo{year}{2020}) \bibinfo{pages}{775--779}. \URLprefix \url{https://www.sciencedirect.com/science/article/pii/S2212827120313238}. \DOIprefix\doi{https://doi.org/10.1016/j.procir.2020.09.135}, \bibinfo{note}{11th CIRP Conference on Photonic Technologies [LANE 2020]}.
\bibitem[{Simonds et~al.(2021)Simonds, Tanner, Artusio-Glimpse, Williams, Parab, Zhao, and Sun}]{simonds2021causal}
\bibinfo{author}{B.~J. Simonds}, \bibinfo{author}{J.~Tanner}, \bibinfo{author}{A.~Artusio-Glimpse}, \bibinfo{author}{P.~A. Williams}, \bibinfo{author}{N.~Parab}, \bibinfo{author}{C.~Zhao}, \bibinfo{author}{T.~Sun},
\newblock \bibinfo{title}{The causal relationship between melt pool geometry and energy absorption measured in real time during laser-based manufacturing},
\newblock \bibinfo{journal}{Applied Materials Today} \bibinfo{volume}{23} (\bibinfo{year}{2021}) \bibinfo{pages}{101049}.

\end{thebibliography}
\end{document}